\documentclass[acmlarge, screen]{acmart}
\usepackage{subfigure}
\usepackage{multirow}
\usepackage{algorithm}
\usepackage{algorithmic}
\usepackage{enumerate}
\usepackage{enumitem}
\usepackage{pdfpages}

\AtBeginDocument{%
  \providecommand\BibTeX{{%
    \normalfont B\kern-0.5em{\scshape i\kern-0.25em b}\kern-0.8em\TeX}}}

\copyrightyear{2020}
\acmYear{2020}
\setcopyright{acmcopyright}

\begin{document}

\title{Enabling Cost-Effective Population Health Monitoring By Exploiting Spatiotemporal Correlation: An Empirical Study}

\author{Dawei Chen}
\authornote{This work was done while Dawei Chen was an academic visitor at the Lancaster University.}
\email{dwchen@mail.ustc.edu.cn}
\affiliation{%
  \institution{Lancaster University}
  \city{Lancaster}
  \country{UK}
}

\author{Jiangtao Wang}
\authornote{Corresponding author}
\email{jiangtao.wang@lancaster.ac.uk}
\affiliation{%
  \institution{Lancaster University}
  \city{Lancaster}
  \country{UK}
}

\author{Wenjie Ruan}
\email{wenjie.ruan@lancaster.ac.uk}
\affiliation{%
  \institution{Lancaster University}
  \city{Lancaster}
  \country{UK}
}

\author{Qiang Ni}
\email{q.ni@lancaster.ac.uk}
\affiliation{%
  \institution{Lancaster University}
  \city{Lancaster}
  \country{UK}
}

\author{Sumi Helal}
\email{s.helal@lancaster.ac.uk}
\affiliation{
  \institution{Lancaster University}
  \city{Lancaster}
  \country{UK}
}


\begin{abstract}
Because of its important role in health policy-shaping, population health monitoring (PHM) is considered a fundamental block for public health services. However, traditional public health data collection approaches, such as clinic-visit-based data integration or health surveys, could be very costly and time-consuming. To address this challenge, this paper proposes a cost-effective approach called Compressive Population Health (CPH), where a subset of a given area is selected in terms of regions within the area for data collection in the traditional way, while leveraging inherent spatial correlations of neighboring regions to perform data inference for the rest of the area. By alternating selected regions longitudinally, this approach can validate and correct previously assessed spatial correlations. To verify whether the idea of CPH is feasible, we conduct an in-depth study based on spatiotemporal morbidity rates of chronic diseases in more than {\em 500} regions around London for over ten years. We introduce our CPH approach and present three extensive analytical studies. The first confirms that significant spatiotemporal correlations do exist. In the second study, by deploying multiple state-of-the-art data recovery algorithms, we verify that these spatiotemporal correlations can be leveraged to do data inference accurately using only a small number of samples. Finally, we compare different methods for region selection for traditional data collection and show how such methods can further reduce the overall cost while maintaining high PHM quality.
\end{abstract}


\maketitle

\section{Introduction}
Due to societal behavior’s changes and ageing population, many chronic and malignant diseases (e.g., heart disease, diabetes, and cancer) are extremely prevalent in our society. According to the World Health Organization (WHO), it has been predicted that, by 2020, chronic diseases will account for almost three-quarters of all deaths worldwide\footnote{\url{https://www.who.int/nutrition/topics/2_background/en/}}. Understanding the changes in population health patterns and trends is important for the planning, monitoring, and evaluation of health programs and policies. To achieve these goals, population health monitoring (PHM) (also called population health surveillance, or public health monitoring) is an extremely important and fundamental block for a nation’s public health system (e.g., UK National Health Services)~\cite{Verschuuren18}. Because of its important role in the health policy-making process, PHM is listed by the WHO as the first of ten essential public health operations~\cite{Verschuuren18}.

There are commonly two ways for healthcare authorities to perform data collection in PHM, that is, integrating digital records of clinic visits~\cite{Perlman17} or conducting surveys among a sample of residents~\cite{Kind98}:
(1) {\itshape Clinic-Visit Data Integration:} In this way, the healthcare authorities need to conduct a huge amount of data integration and linkage from multiple information systems to get an overview picture (e.g., the morbidity rate of heart attack in different towns in Northwest England). However, the data integration is non-trivial due to the following reasons. First, the data access may be a barrier as some of data entries are privacy sensitive. For example, to know the population health statistics in a city, we need to access individual-level sensitive health data. Thus, the engineers must do substantial extra work in data privacy protection (e.g., data anonymity operations) before starting the data integration. Second, the data structure and database design might be different for heterogeneous systems from multiple clinics or hospitals, thus increasing the difficulty and cost of data linkage.
(2) {\itshape Survey-Based data collection:} In this way, the administrators need to recruit a representative group of residents and collect data via interviews or self-managed questionnaires. In order to minimize the bias, the number of samples should be large enough with appropriate demographic distributions. Therefore, this process is time-consuming and incurs high cost including labor cost for administrators and incentive payment for survey participants. In summary, both of these traditional data collection approaches are of high cost and time-consuming, thus novel and more cost-effective solution of data collection for PHM is urgently needed.

In this paper, we introduce a cost-effective approach to PHM which we call Compressive Population Health (CPH), and evaluate its feasibility through three extensive analytical studies. In this approach, a coverage area is divided into a number of regions of which a subset is selected for data collection in the traditional way, while the other regions’ missing data is “recovered” by leveraging the inherent spatial correlations with their selected neighboring regions. By alternating and swapping regions longitudinally as the CPH system collects a fresh set of data, prior missing data regions turn into selected regions, validating and correcting the prior correlation assessment while keeping the same cost-effectiveness as CPH progresses. We have to emphasize that this is NOT the style of work aiming to develop new computer science algorithms/methods. Instead, its contribution lies in a novel solution to an important public health problem with its feasibility investigations enabling by computer science. This is the first work systematically exploring if the spatiotemporal correlations can be leveraged for accurate population health data inference based on a small number of samples by answering three progressive research questions. 

The rest of the paper is organized as follows: Section 2 presents the Compressive Population Health idea and articulates its novelty and expected impact. Section 3 describes the datasets and the analysis of spatiotemporal correlations. Section 4 introduces the missing data entry recovery algorithms for CPH and compares their performance under various settings. Section 5 proposes two TS-A selection methods and compares their effectiveness. Section 6 reviews the related works from different perspectives. In Section 7, we summarize the findings and discuss the implications with future work directions. Section 8 concludes this study.

\section{Compressive Population Health}
\subsection{Basic Idea of CPH}
Compressive sensing~\cite{Baraniuk07} is a signal processing technique for efficiently acquiring and reconstructing a signal, which has been successfully applied in many domains such as computer vision, sensor networks, and urban sensing~\cite{Kong13,Quer12,Guestrin05,Zhu12,Leye15}. Compressive sensing is based on the principle that, through optimization, the sparsity of a signal can be exploited to recover it from far fewer samples than required. With the challenge of high cost for PHM in mind and inspired by the idea of compressive sensing, we propose a potentially disruptive paradigm for data collection of PHM, which we name Compressive Population Health (CPH). CPH only selects a subset of regions (called Traditionally Sensed Areas, TS-A for short) to conduct traditional data collection (i.e., either the clinic-visit-based data integration or survey-based approach). For the rest of un-collected regions (called “Inferred Areas”, IF-A for short), CPH uses the data collected from TS-A and leverages the inherent data correlations to do inference.
\begin{figure}
    \centering
    \includegraphics[width=0.6\linewidth]{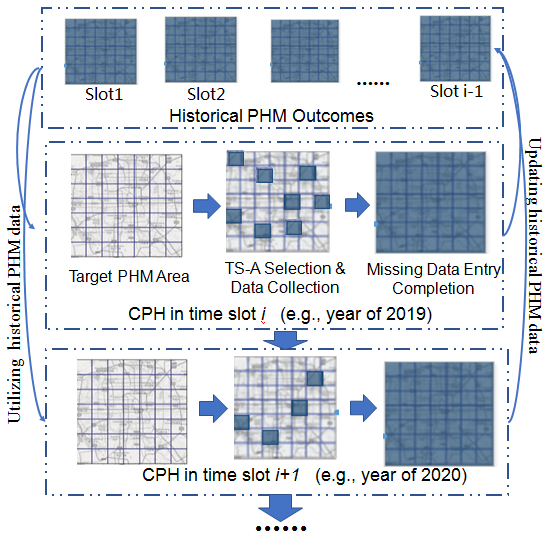}
    \caption{An illustration of the basic idea of CPH}
    \label{fig:CPH}
\end{figure}

Figure~\ref{fig:CPH} briefly illustrates the basic process of CPH. For a given targeting PHM area (e.g., south England) divided into several regions (e.g. ward-level ones), a given monitoring time slot (e.g., the year of 2019), and the historical PHM outcomes in previous years (e.g., morbidity rate of multiple chronic diseases from 2010 to 2018), CPH will work based on the following workflow. First, it utilizes the latent spatiotemporal correlations embedded in the historical monitoring outcomes to select a number of most informative regions as TS-A based on the budget constraints. Then, healthcare administrators will perform traditional data collection practice (e.g., conducting surveys or integrating clinic-visit record) to obtain the population health outcome of TS-A. Third, CPH deploys missing data entry recovery algorithms to accurately infer the data in IF-A so as to form complete PHM results. Finally, the completed data will be used to update the historical data for the CPH of future time slots (e.g., the year of 2020). In summary, we expect that through CPH, healthcare administrators can complete the PHM tasks with a significantly reduced cost.

\subsection{Research Questions and Contributions}
Although the idea of CPH intuitively seems to be promising with potential in significantly reducing the cost of monitoring health and well-being, we need to rigorously investigate whether it is practically feasible, which is the primary objective of this study. Specifically, this paper aims to address the following three research questions:
\begin{itemize}[leftmargin=*]
    \item {\itshape\textbf{RQ1: Does spatiotemporal correlations do exist for PHM?}}
    The foundation of the success of compressive sensing in sensor-network-based environment monitoring is the existence of strong spatiotemporal correlations for environmental data (e.g., air quality and temperature). With the spatiotemporal correlation, we can only do the sensing in a subset of geographical grids or time slots, and then infer the remaining. Therefore, in order to study the feasibility of CPH, we need to first check whether similar correlations also exist for population health outcomes.
    \item {\itshape\textbf{RQ2:} If the answer to RQ1 is YES, then \textbf{are the correlations significant enough to support the missing data inferences for IF-A?}}
    In other words, if such correlations do exist, we would like to confirm whether the state-of-the-art missing data recovery algorithms can fully utilize these correlations to achieve accurate data inference for IF-A. 
    \item {\itshape\textbf{RQ3:} If the answers to both RQ1 and RQ2 are YES, \textbf{does the selection of TS-A matter in data collection of CPH?}}
    For example, in the problem of compressive sensing in sensor networks, selecting appropriate places for sensor placement can further reduce the cost or improve the inference quality. Similarly, we attempt to study whether the TS-A selection plays a similarly crucial role in the process of CPH.
\end{itemize}
To address the above research questions, this paper conducts an in-depth study based on a dataset with morbidity rates of chronic diseases data in more than 500 regions (called wards) in London for over 10 years. Through a data-driven study from different perspectives, the results first reveal the existence of spatiotemporal correlations for the evolution of chronic diseases (for \textbf{addressing RQ1}), and then demonstrate the implications for such correlations in our proposed CPH solution (for \textbf{addressing RQ2 and RQ3}). The contributions of this paper lie in the novel solution to an important public health problem and the investigations of its feasibility with the state-of-the-art computational techniques. Specifically, the contributions in this paper can be summarized as follows:

First, we perform a descriptive analysis of the relationship between the morbidity rate difference and spatiotemporal factors of multiple chronic diseases. The results of analysis reveal that the difference of morbidity rate is generally significantly correlated with factors such as inter-region distances and time differences (\textbf{section 3, for addressing RQ1}). Besides, we also present some possible explanations supported by existing research evidences in the domain of public health.

Second, we formally formulate the data inference in CPH as the problem of missing data entry completion, and then deploy a set of state-of-the-art missing data recovery methods to solve it under various settings. The results indicate that with appropriate missing data inference algorithms and collected population health data from a relatively small number of TS-A, the data inference for IF-A can be highly accurate (\textbf{Section 4, for addressing RQ2}).

Third, we adopt two methods to select the most informative regions as TS-A, and then execute various missing data recovery algorithms for performance comparison with the random selection. The comparison results demonstrate that these two methods significantly outperform the random selection, which suggests that the optimized selection of TS-A can significantly improve the implementation of CPH (\textbf{section 5, for addressing RQ3}).
 
\section{Spatiotemporal Correlation Analysis}


\subsection{Description of Datasets}

In this study, we use three datasets which can be 
collected from UK government's website without any licences:

\begin{itemize}[leftmargin=*]
    \item {\itshape Chronic disease prevalence dataset:} Since 1 April 2004, The Quality and Outcomes Framework (QOF) was introduced as part of the General Medical Services (GMS) by The National Health Service (NHS)
    \footnote{\url{https://digital.nhs.uk/data-and-information/publications/statistical/quality-and-outcomes-framework-achievement-prevalence-and-exceptions-data}} which aims to improve the quality of care patients by rewarding practices for the quality of care they provide to their patients. NHS annually publishes healthy data (including 17 chronic diseases, see Table~\ref{tab:disease}) ranging from 1 April to next 31 March which covers up to 94.8 percent of all general practices. For each type of disease, these data include the morbidity rate as an indicator of disease prevalence to represent the ratio of the number of patients on each register to the number of all patients on the practice list. We have collected ten-year data from 2009 to 2018. In this paper, we use dataset ``2009'' to represent the annual chronic disease prevalence from April 2008 to March 2009.
    
    \item {\itshape Dataset of Ward Boundaries of London:} This dataset\footnote{It is collected from the website \url{http://data.ordnancesurvey.co.uk}} is collected from Great Britain's national mapping agency which provides the most accurate and up-to-date geographic data for government, business and individuals, it contains the fine-grained information of ward\footnote{Ward is the electoral district at sub-national level.} boundaries in London. Specifically, the dataset includes names, shapes and codes of 630 wards in London, as shown in Figure~\ref{fig:map in London}.
    
    \item {\itshape Dataset of Geographic Information in London:} This dataset includes the longitudes and latitudes of wards in London\footnote{It is collected from the website \url{https://geoportal.statistics.gov.uk/}}. Those geographic coordinates will be used to calculate Euclidean distances between wards for an in-depth spatial correlation analysis.
    
\end{itemize}

\begin{table}
\centering
  \begin{tabular}{cc}
  \hline
    Disease name&Abbreviation\\
    \hline
    Atrial Fibrillation & AF 
    \\ Coronary Heart Disease & CHD
    \\Heart Failure & HF
    \\Hypertension & HYP
    \\Stroke and Transient Ischaemic Attack & STIA
    \\Peripheral Arterial Disease & PAD 
    \\Asthma &AST    
	\\Chronic Obstructive Pulmonary Disease & COPD
	\\Obesity & OB 
	\\Cancer & CAN 
	\\ Chronic Kidney Disease & CKD
	\\ Diabetes Mellitus & DM
	\\ Palliative Care$^{\rm a}$ & PC 
	\\Dementia & DEM
	\\ Depression & DEP
	\\ Epilepsy & EP
	\\ Learning Disabilities &LD
	\\ Mental Health & MH \\ \hline
\end{tabular}
  \caption{Chronic diseases' name and its abbreviations}
  \label{tab:disease}
\end{table}

\subsection{Spatial Correlation Analysis}
\begin{figure}[htb]
    \centering
    \includegraphics[width=0.6\linewidth]{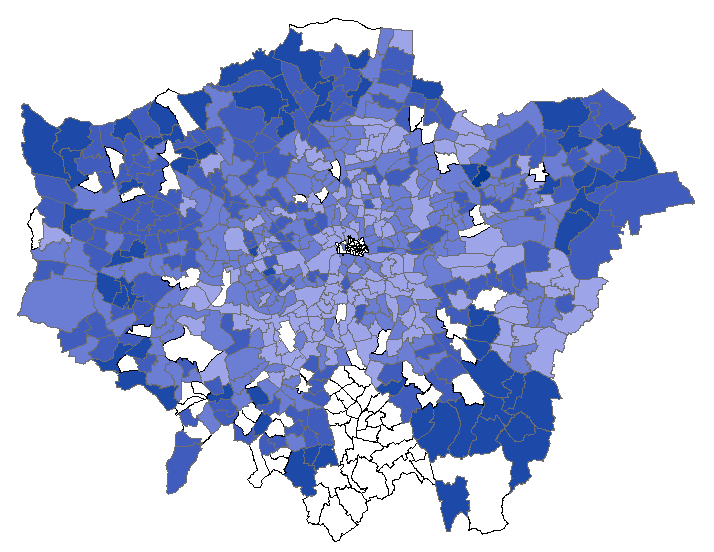}
    \caption{The ward-level boundary line of London and ward-grained morbidity rate distribution of CHD (darker colors indicate higher morbidity rate, blank parts represent missing data)}
    \label{fig:map in London}
\end{figure}

This section will present our analysis results on spatial correlation of chronic morbidities. From Figure~\ref{fig:map in London}, we indeed observe some spatial correlations among morbidities. We can see that the disease morbidity rates in adjacent wards generally reveal a certain similarity. However, sometimes they also change over location significantly and non-linearly, and morbidity rate with a shorter distance may not always be more similar than those with a farther distance.

To quantify the spatial similarity, we first calculate the Euclidean distances of all ward pairs. Assuming that we have $N$ wards, i.e., $W=(r_1,\cdots,r_i,\cdots,r_N)$, then we can get a set of ward-pairs $P = \{p_{i,j}=(r_i,r_j) \mid i>j, r_i,r_j \in W\}$ which excludes the pairs consisted by the same wards, where $p_{i,j}$ represents a ward-pair consisted by ward $r_i$ and $r_j$. Then based on $p_{i,j}$, we categorise these pairs into $M$ groups denoted by $G=\{g_1, g_2, \cdots, g_{M}\}$. In this paper, we set $M=53$ since the distances of ward-pairs in London range from 0.5km to 53km. In this way, group $g_i$ contains all the ward-pairs with distance between $(i-1)$km and $i$km.
%
For example, the 1st group $g_1$ contains all the ward-pairs whose distances are less than 1 km, while the group $g_2$ contains ward-pairs whose distances are between 1 km and 2 km. For a certain group with $m$ ward-pairs, we use $X = (x_1,x_2,\cdots,x_m)$ to include all the first wards in $m$ ward-pairs, and $Y = (y_1,y_2,\cdots,y_m)$ to include all the second wards in all $m$ ward-pairs. Then for a certain disease, we use $A=(a_1,a_2,\cdots,a_m)$ and $B=(b_1,b_2,\cdots,b_m)$ to represent the morbidities of the wards in $X$ and $Y$. In this paper, we adopt four difference indicators, including Arithmetic Difference (AD), Euclidean Distance (ED), Pearson Distance (PD) and Cumulative Distance of Dynamic Time Warping (CDDTW), to quantitatively measure the spatial correlation.

\begin{itemize}[leftmargin=*]
    \item Arithmetic Difference (AD): it reflects the arithmetic difference of diseases' morbidity. 
    \begin{equation}
        AD(A,B)=\frac{\sum_{i=1}^{m}{\lvert a_i-b_i\rvert}}{m}
        \label{Arithmetic_Difference}
    \end{equation}
    
    \item Euclidean Distance (ED): it is usually used to reflect the difference in numerical value of two vectors, so we only use it to show the absolute similarity of morbidity rates between two wards.
    \begin{equation}
        ED(A,B) = \frac{\sqrt{\sum_{i=1}^{m}{(a_i-b_i)^2}}}{m}
        \label{Eular_Distance}
    \end{equation}
    \item Cumulative Distance of Dynamic Time Warping  (CDDTW): in time series analysis, dynamic time warping (DTW) is one of the algorithms for measuring similarity between two temporal sequences, in which CDDTW is the metric to measure the distance between two time-series~\cite{Keogh2005Exact}. CDDTW improves the inability of ED to deal with local time shifting. It can also eliminate possible time interference caused by the local stretch or compression.$$CDDTW(A,B)=\frac{C(m,m)}{m}$$
    \begin{equation}
        \begin{split}
            C(m,m)=&\sum_{i=1}^m\sum_{j=1}^m \Big(min\big(min(C(i-1,j),(C(i,j-1))),(C(i-1,j-1))\big)\\&+distance(i,j)\Big)
        \end{split}
        \label{DTW_Distance}
    \end{equation}
    where \(distance(i,j) = \sqrt{(a_i-b_j)^2}\).
    \item Pearson Distance (PD): it is used to measure the direction difference of two vectors~\cite{Seiler2010Numerical}, so we use it to show the difference of changing trend of morbidity.
    \begin{equation}
        PD(A,B)=1-\frac{\sum_{i=1}^{m}(a_i-\Bar{A})(b_i-\Bar{B})}{\sqrt{\sum_{i=1}^m (a_i-\Bar{A})^2}\sqrt{\sum_{i=1}^{m}(b_i-\Bar{B})^2}}
        \label{Pearson_Distance}
    \end{equation}
    where $\Bar{A}$ and $\Bar{B}$ represent the mean value of vector \(A\) and \(B\).
\end{itemize}

\begin{figure}[htb]
    \centering
    \subfigure[CHD]{
    \includegraphics[width=0.23\linewidth]{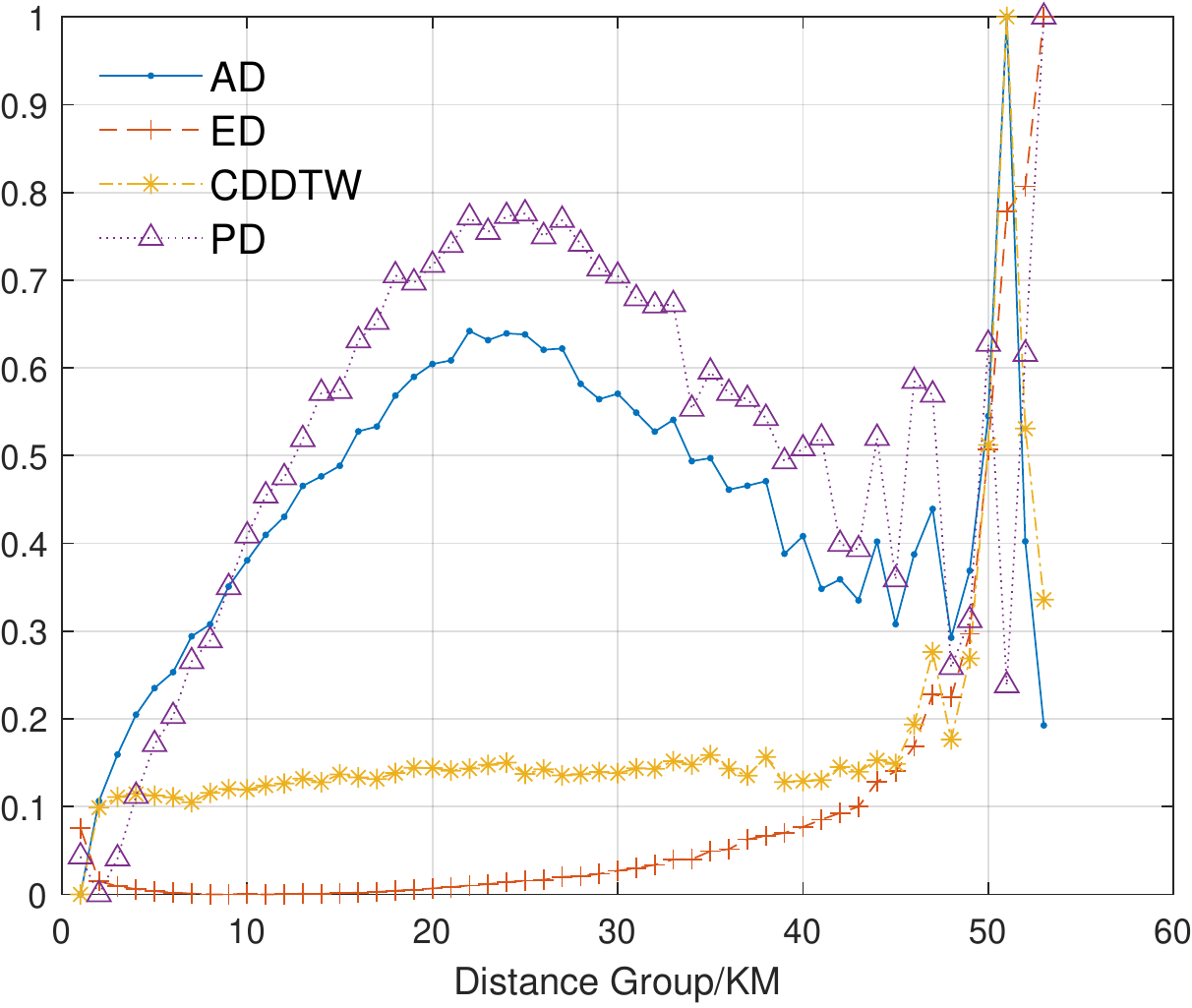}
    \label{spatial correlation fig:subfig:a}}
    \subfigure[HYP]{
    \includegraphics[width=0.23\linewidth]{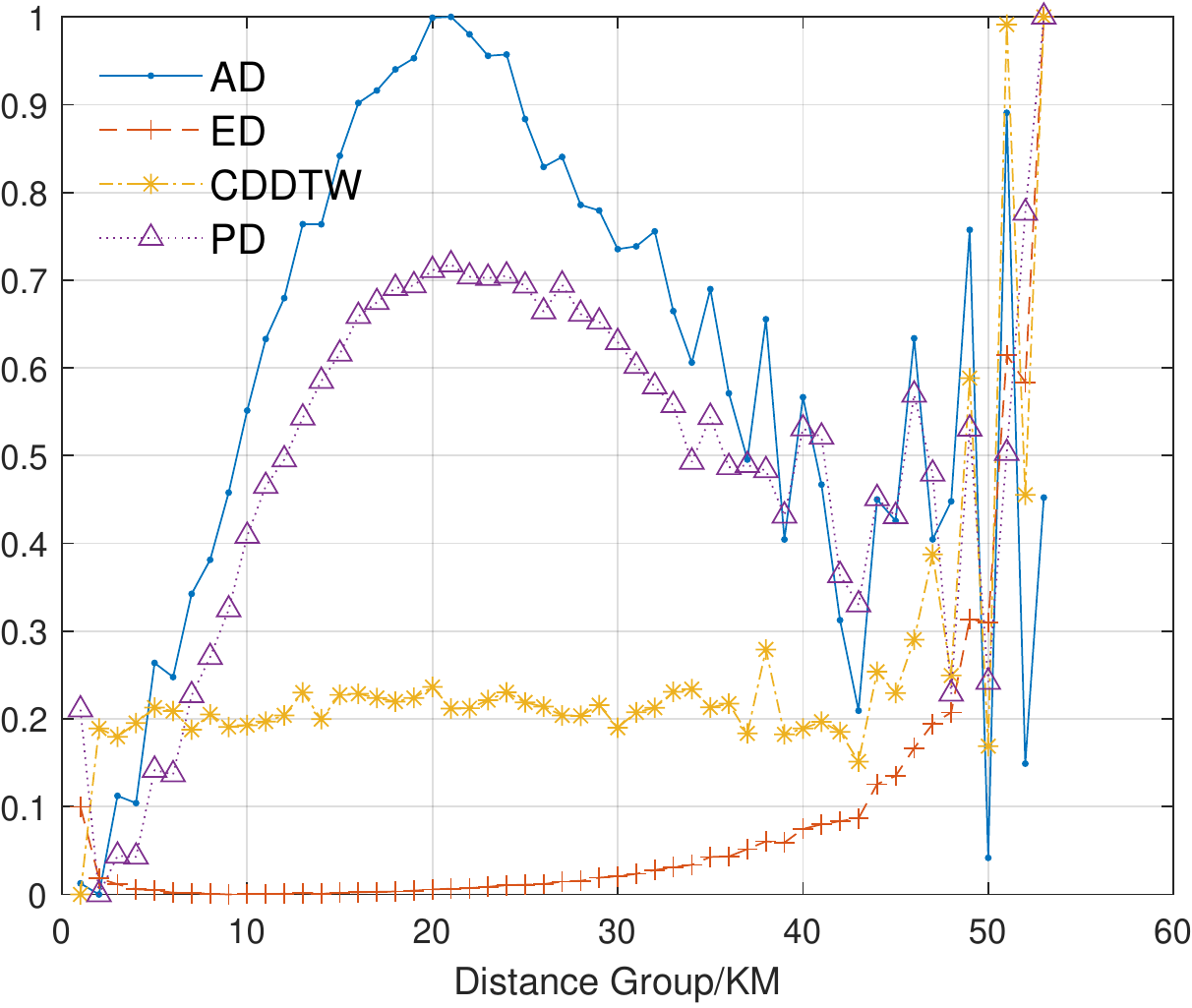}
    \label{spatial correlation fig:subfig:b}}
    \subfigure[STIA]{
    \includegraphics[width=0.23\linewidth]{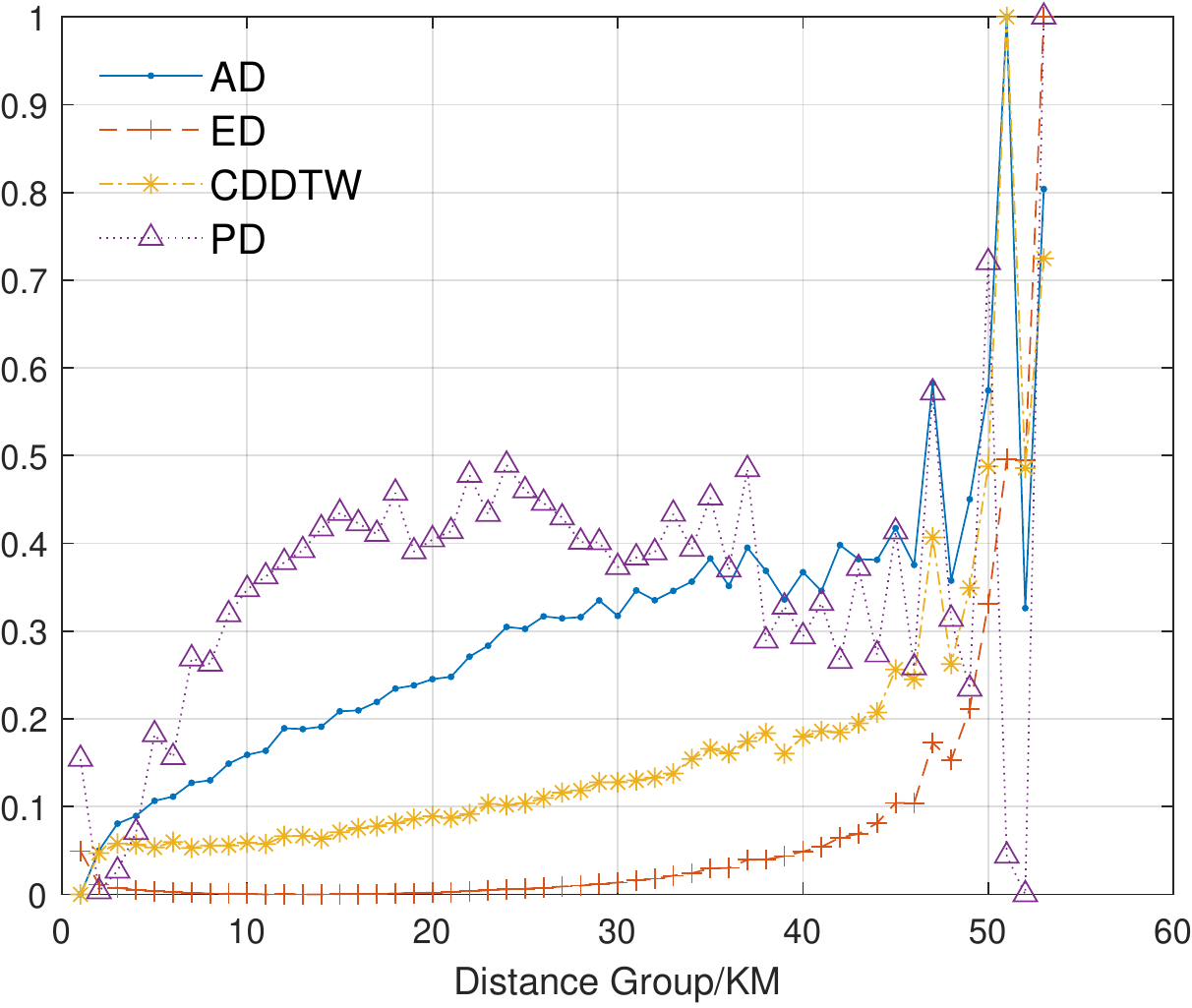}
    \label{spatial correlation fig:subfig:c}}
    \subfigure[CAN]{
    \includegraphics[width=0.23\linewidth]{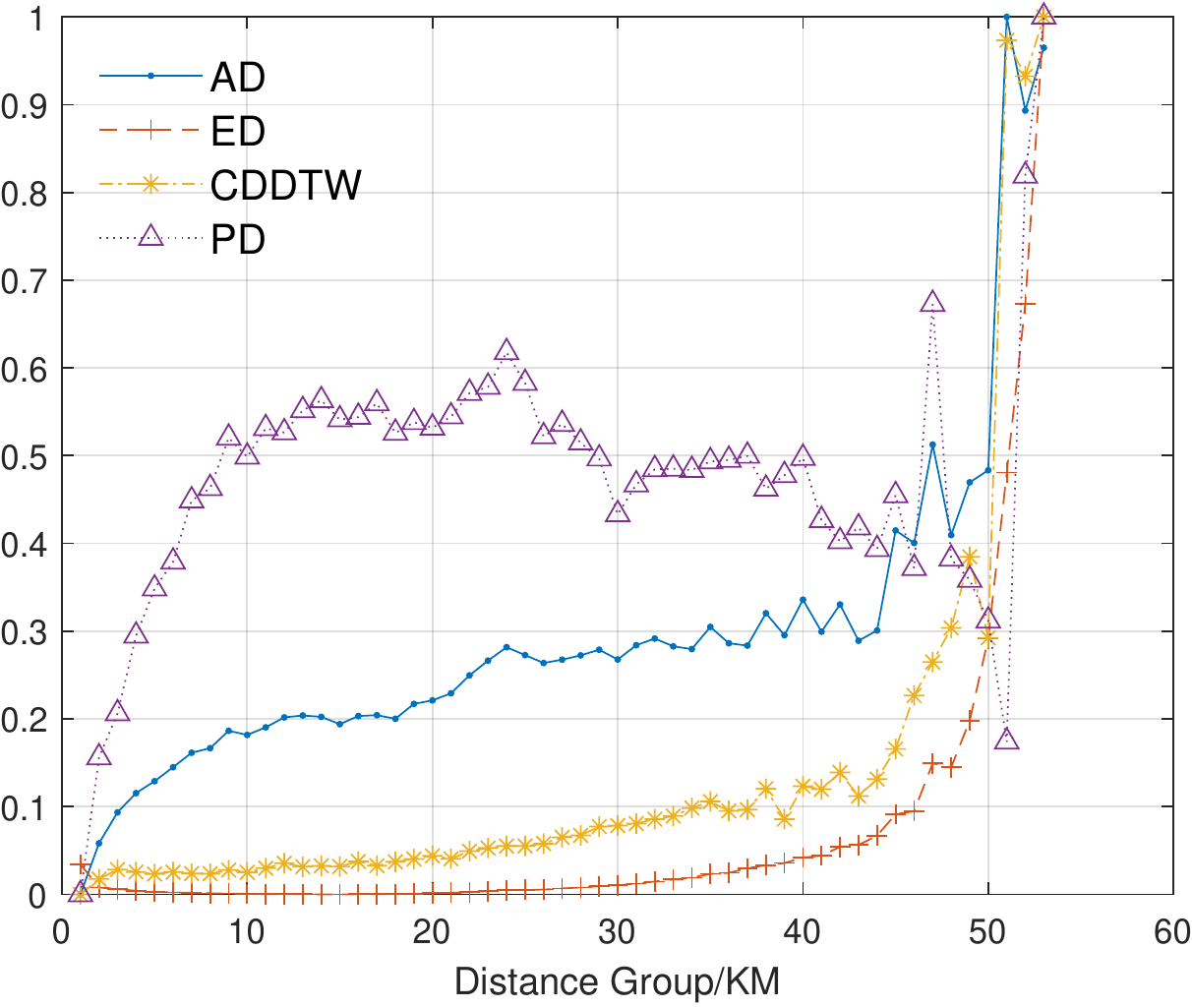}
    \label{spatial correlation fig:subfig:d}}
    \subfigure[EP]{
    \includegraphics[width=0.23\linewidth]{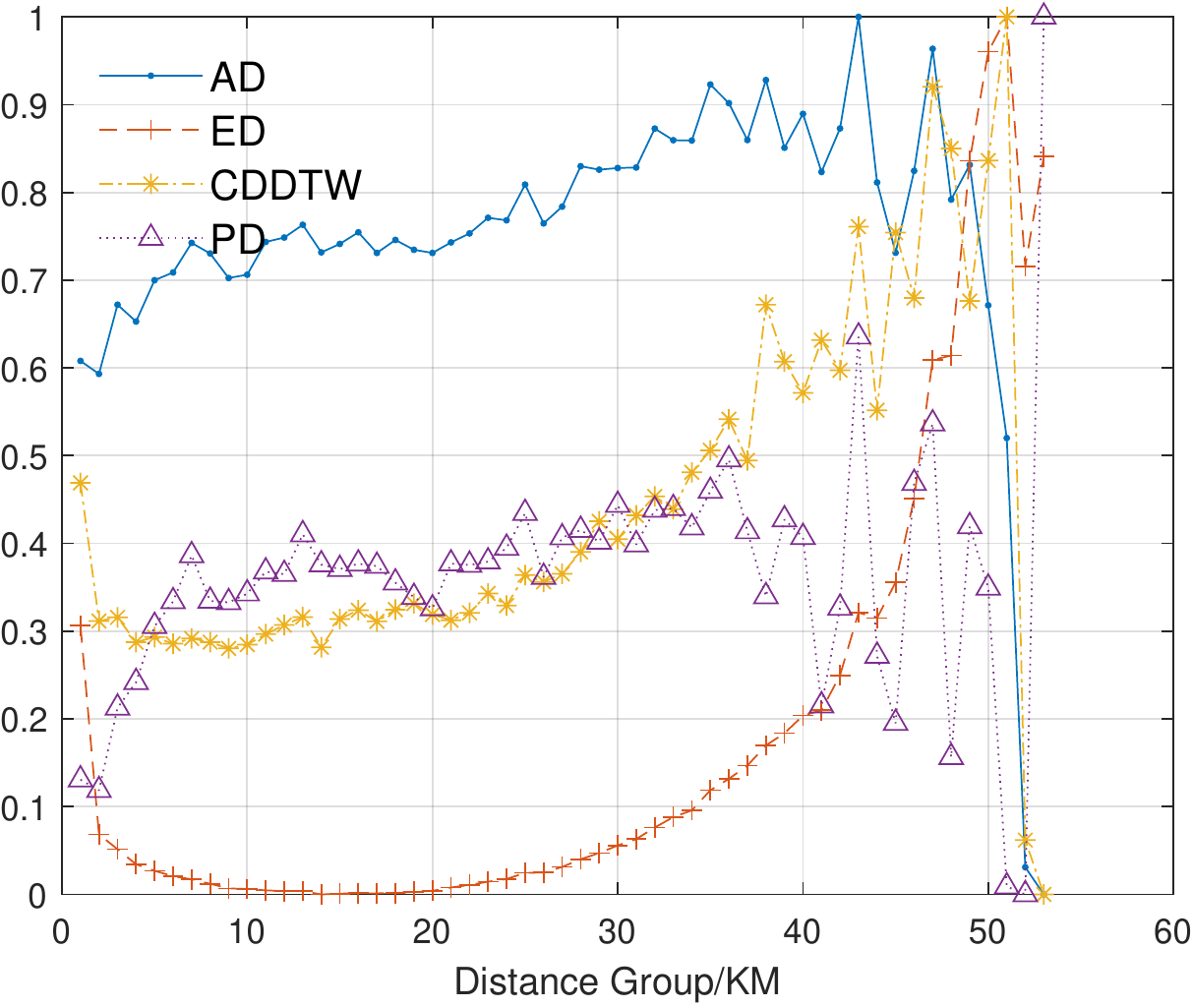}
    \label{spatial correlation fig:subfig:e}}
    \subfigure[HF]{
    \includegraphics[width=0.23\linewidth]{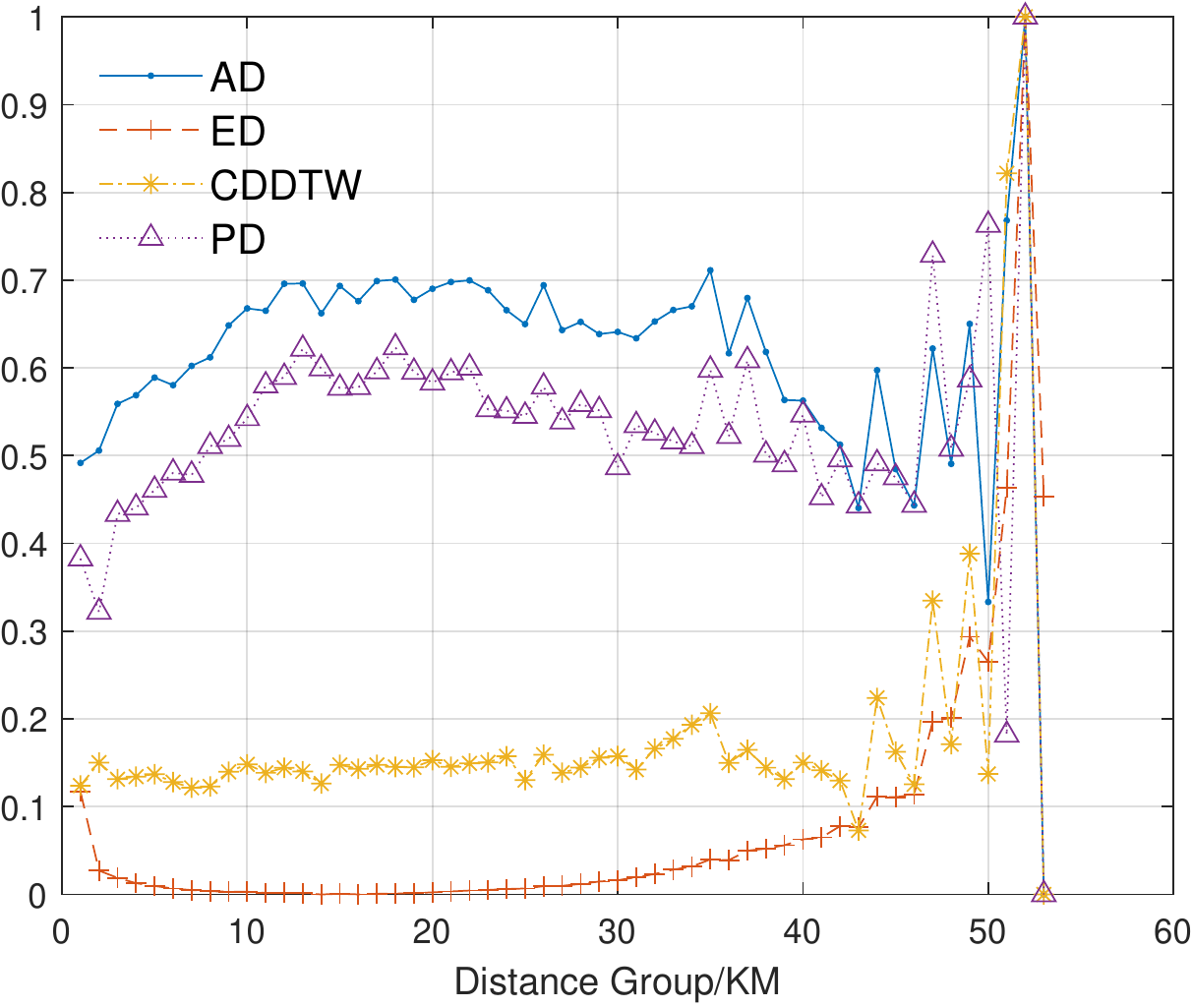}
    \label{spatial correlation fig:subfig:f}}
    \subfigure[PC]{
    \includegraphics[width=0.23\linewidth]{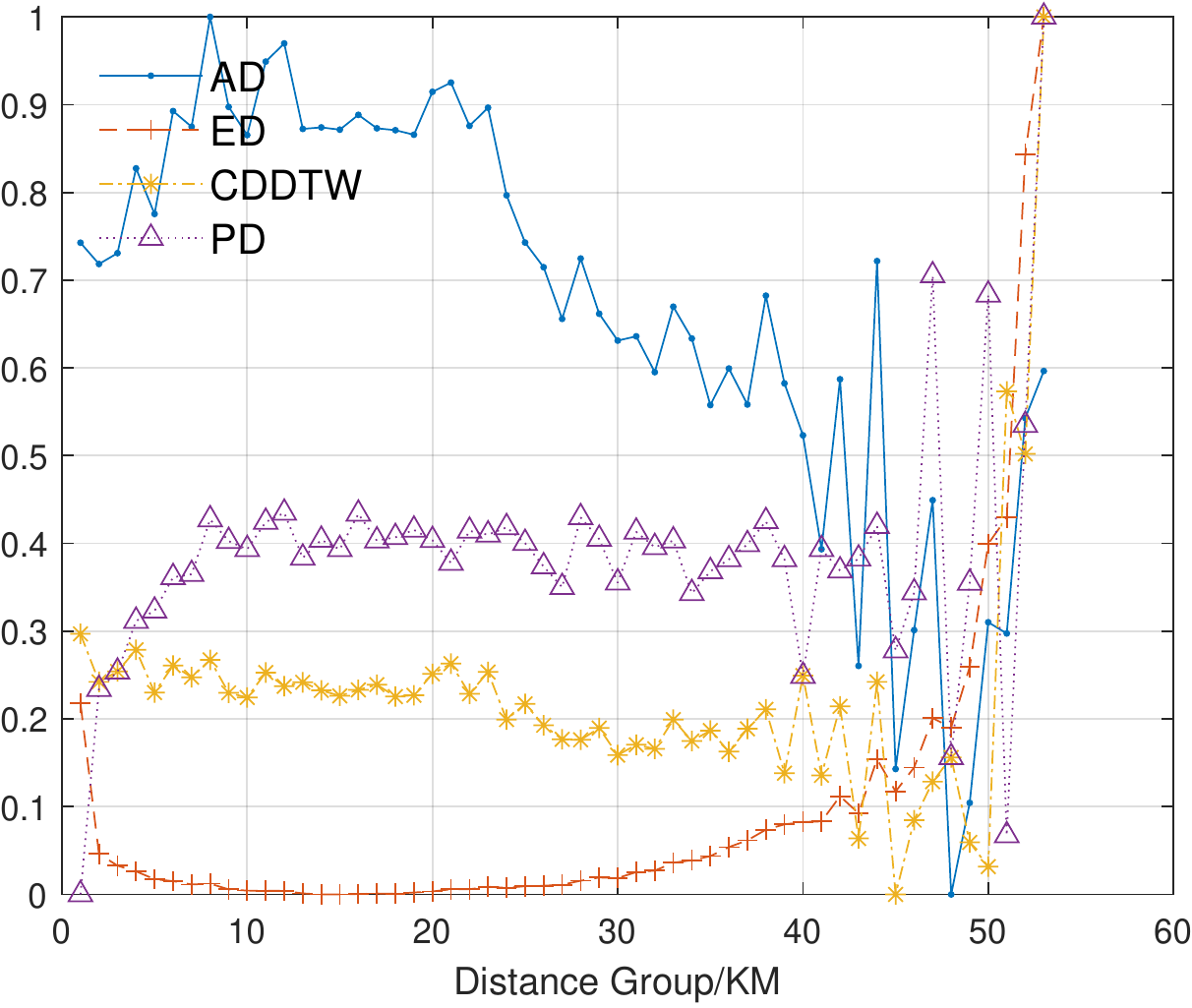}
    \label{spatial correlation fig:subfig:g}}
    \subfigure[DEM]{
    \includegraphics[width=0.23\linewidth]{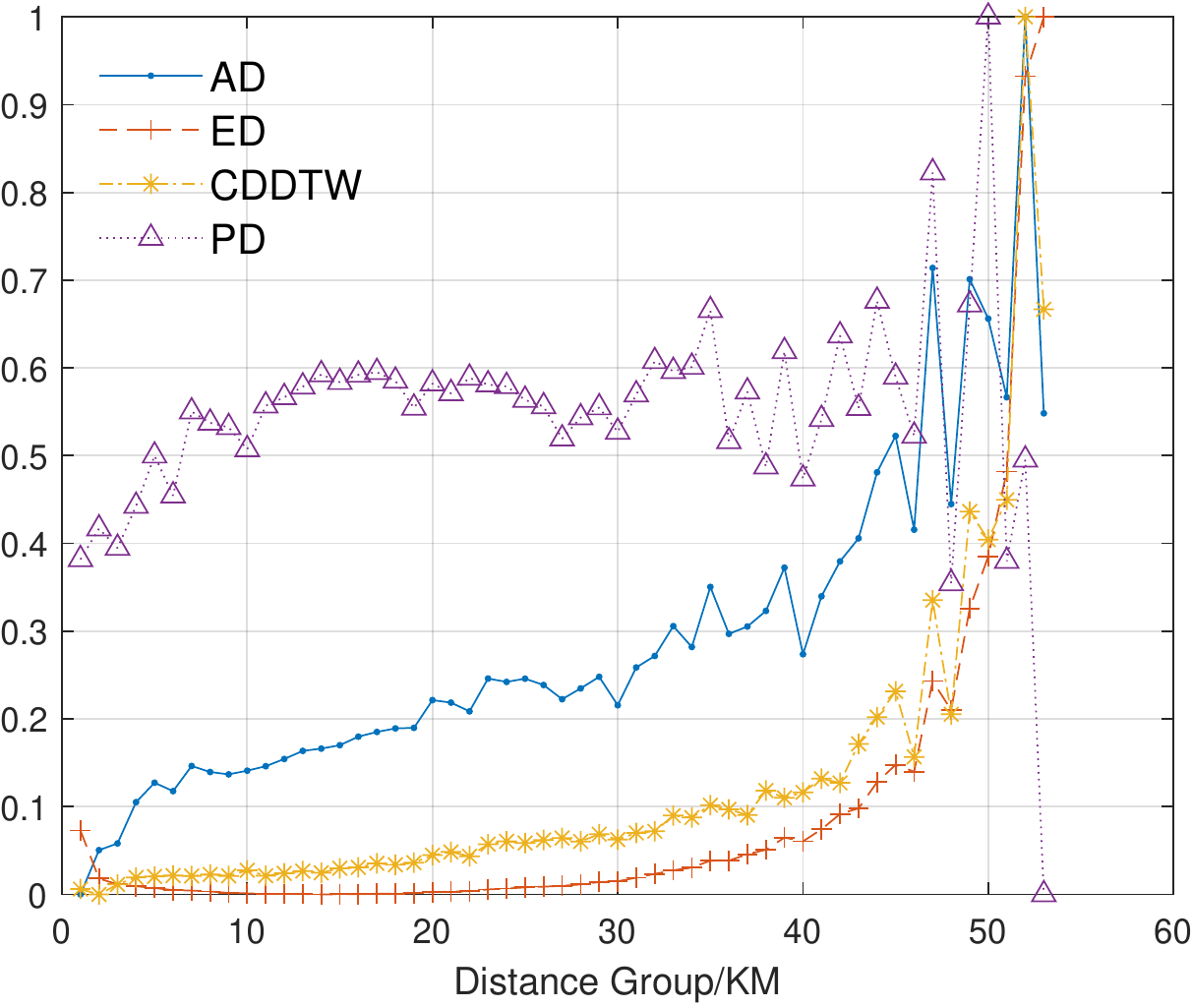}
    \label{spatial correlation fig:subfig:h}}
    \caption{Morbidity difference changes as distance increases}
    \label{fig:spatial correlation}
\end{figure}

Due to the page limitation, we present the empirical results\footnote{The data are normalized in order to fit the figures.} of eight typical diseases in Figure~\ref{fig:spatial correlation}, including CHD, HYP, STIA, CAN, EP, HF, PC and DEM. We can further make the following observations:

\begin{itemize}[leftmargin=*]
    \item The spatial correlations generally {\em exist} within certain geographically scale for most of the diseases. For example, for CHD, HYP, SITA and CAN (see Figure~\ref{spatial correlation fig:subfig:a}, \ref{spatial correlation fig:subfig:b}, \ref{spatial correlation fig:subfig:c} and \ref{spatial correlation fig:subfig:d}),  all four difference indicators are relatively small for the set of ward pairs within distances less than 10 km. Moreover, for STIA, DEM (see Figure~\ref{spatial correlation fig:subfig:c}\ref{spatial correlation fig:subfig:h}), the inter-ward difference of morbidity rate becomes larger as the distance increases within a certain scale.  
    
    \item Although the spatial temporal correlations do exist, they are non-linear and even disappear out of certain geographical scale. For example, from Figure~\ref{spatial correlation fig:subfig:b}, we can observe that the spatial correlations in HYP show strong nonlinear. The ED index first increases and then decreases, and shows strong fluctuates after 40km.
    
    \item The significance of correlations are different among various diseases. For example, diseases such as CAN, DEM, STIA possess stronger correlation than that of EP and PC. 
\end{itemize}


Through these observations, we can conclude that the spatial correlations of morbidity definitely exist especially when the distance of two wards does not exceed a certain threshold. But the spatial correlations are highly nonlinear and complex. The complicity lies in two aspects. One is that spatial correlation does not show linear relation with distance, and only exist within a certain geographic scale. The other is that the spatial correlation varies significantly among different diseases. The existence of spatial correlation potentially brings new opportunities to our proposed CPH but its complexity and non-linearity also bring challenges.

\subsection{Temporal Correlations Analysis}

\begin{figure}[htb]
    \centering
    \subfigure[COPD-AD]{
    \includegraphics[width=0.23\linewidth]{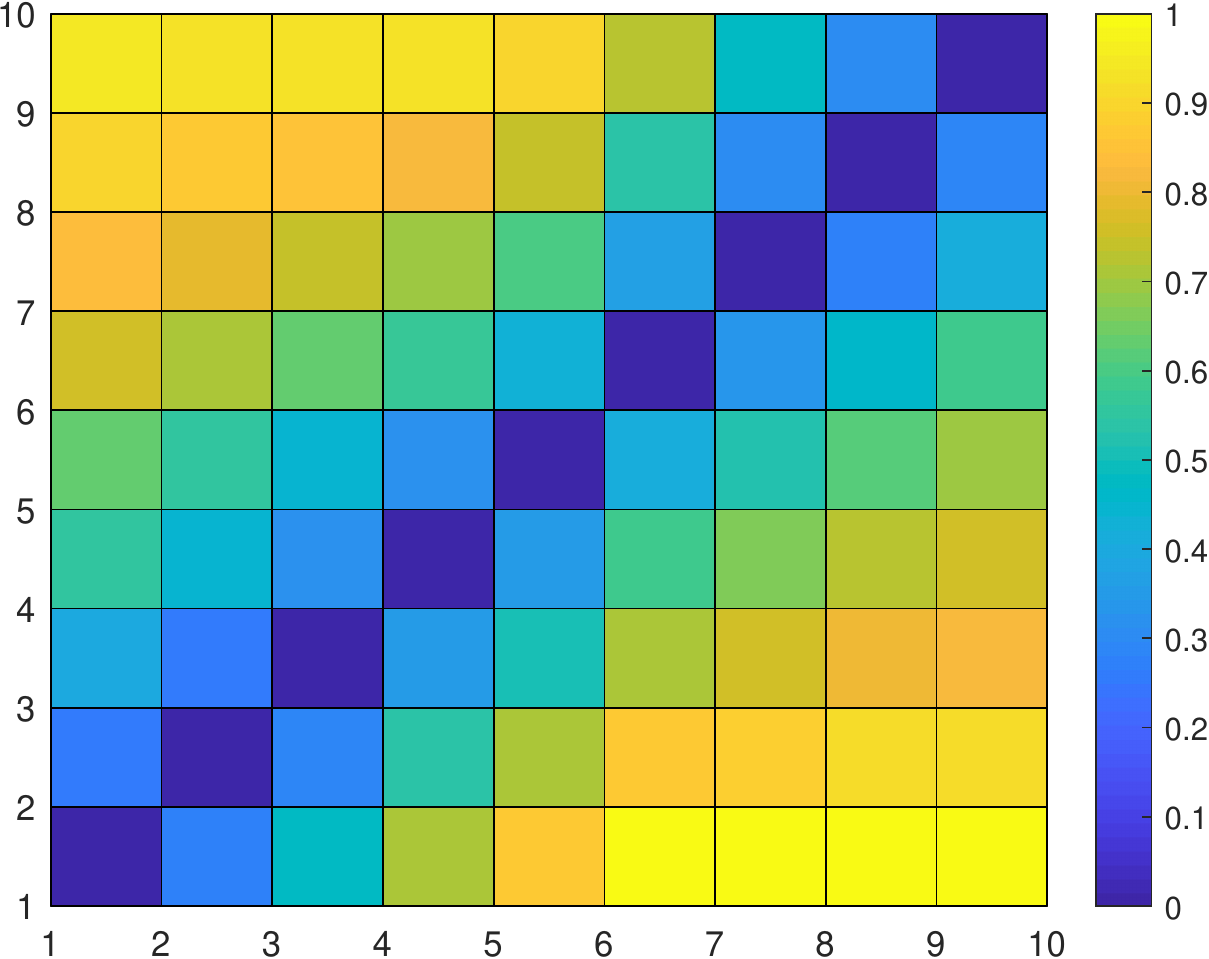}
    \label{fig:temporal correlation a}}
    \subfigure[COPD-ED]{
    \includegraphics[width=0.23\linewidth]{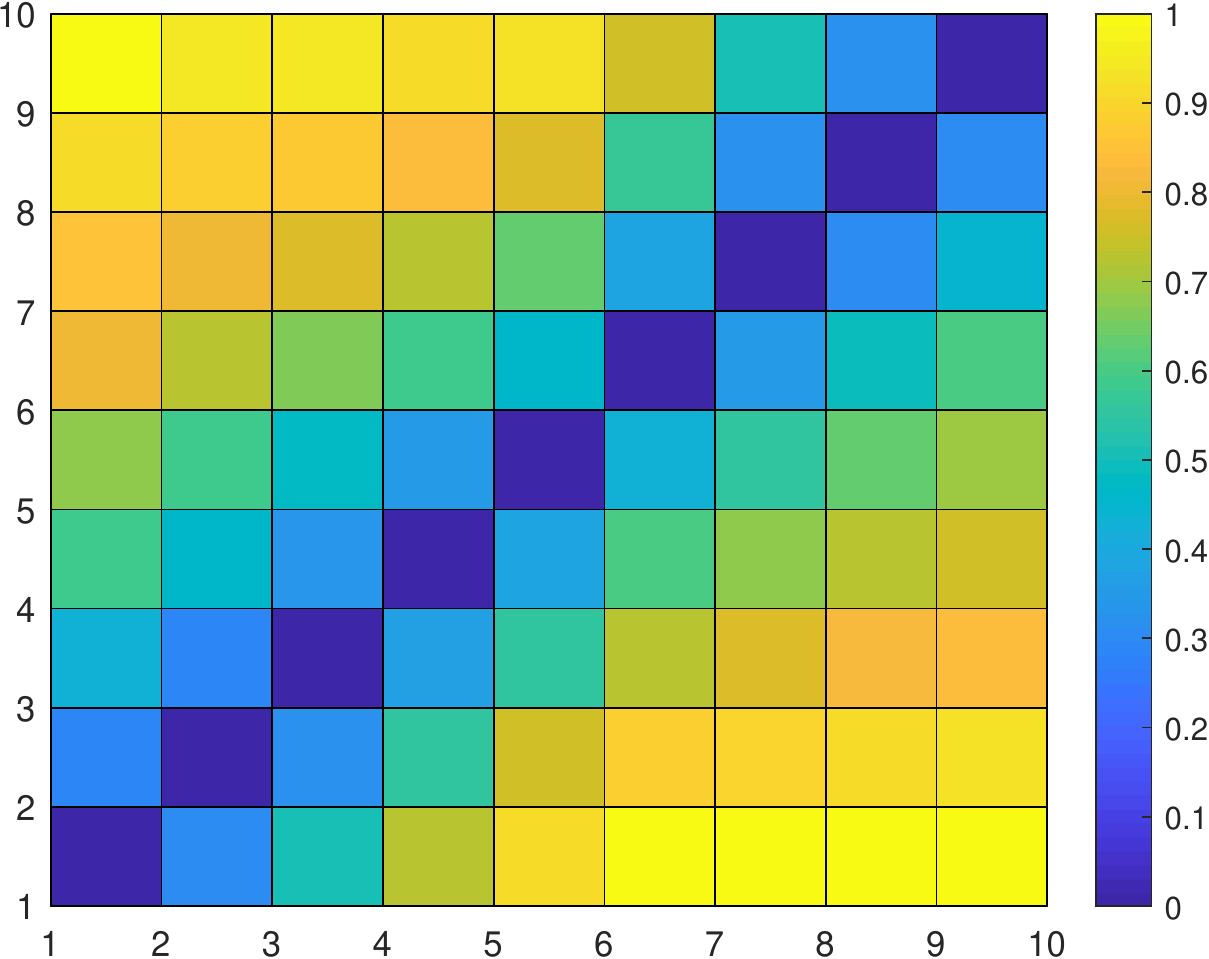}
    \label{fig:temporal correlation b}}
    \subfigure[COPD-CDDTW]{
    \includegraphics[width=0.23\linewidth]{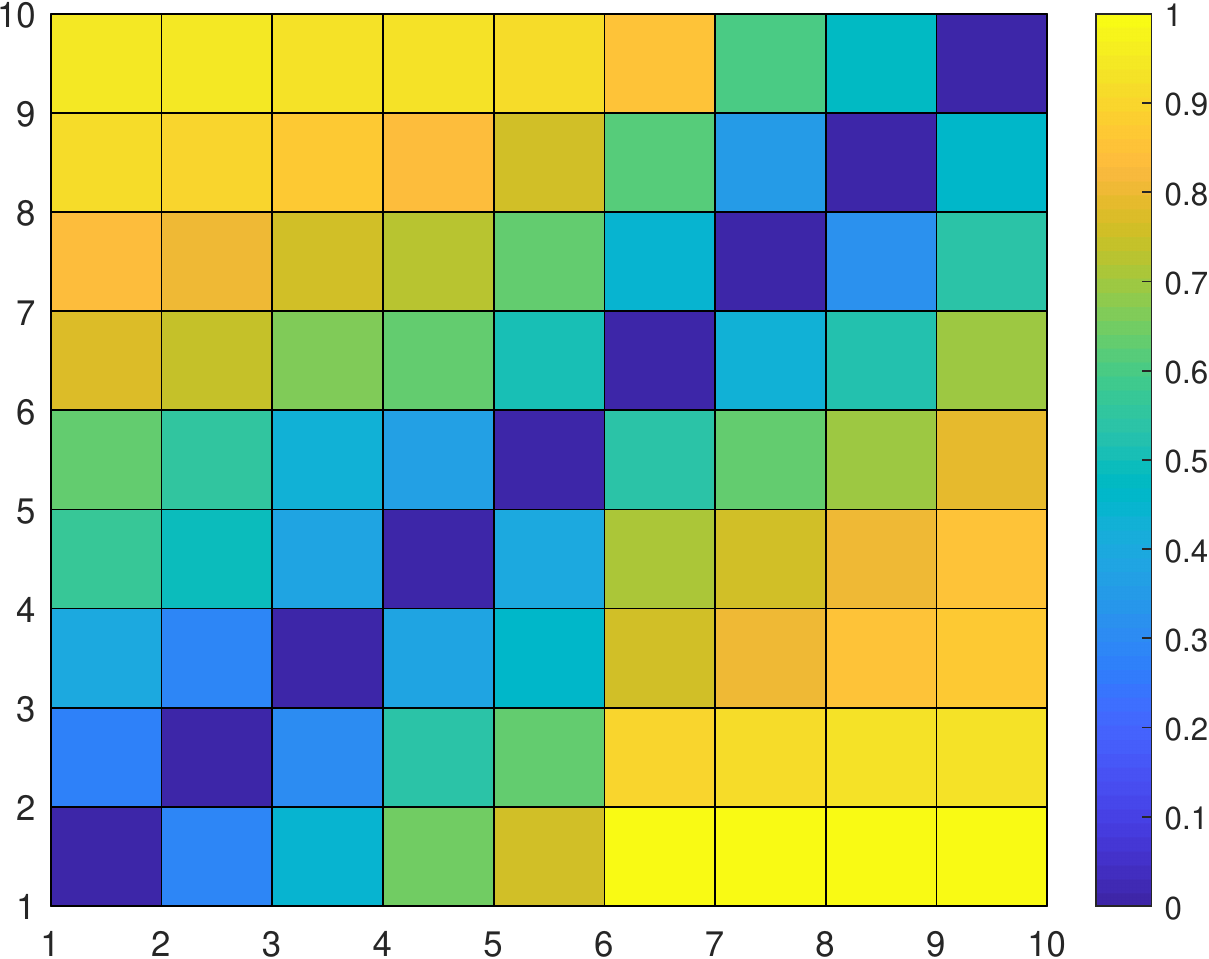}
    \label{fig:temporal correlation c}}
    \subfigure[COPD-PD]{
    \includegraphics[width=0.23\linewidth]{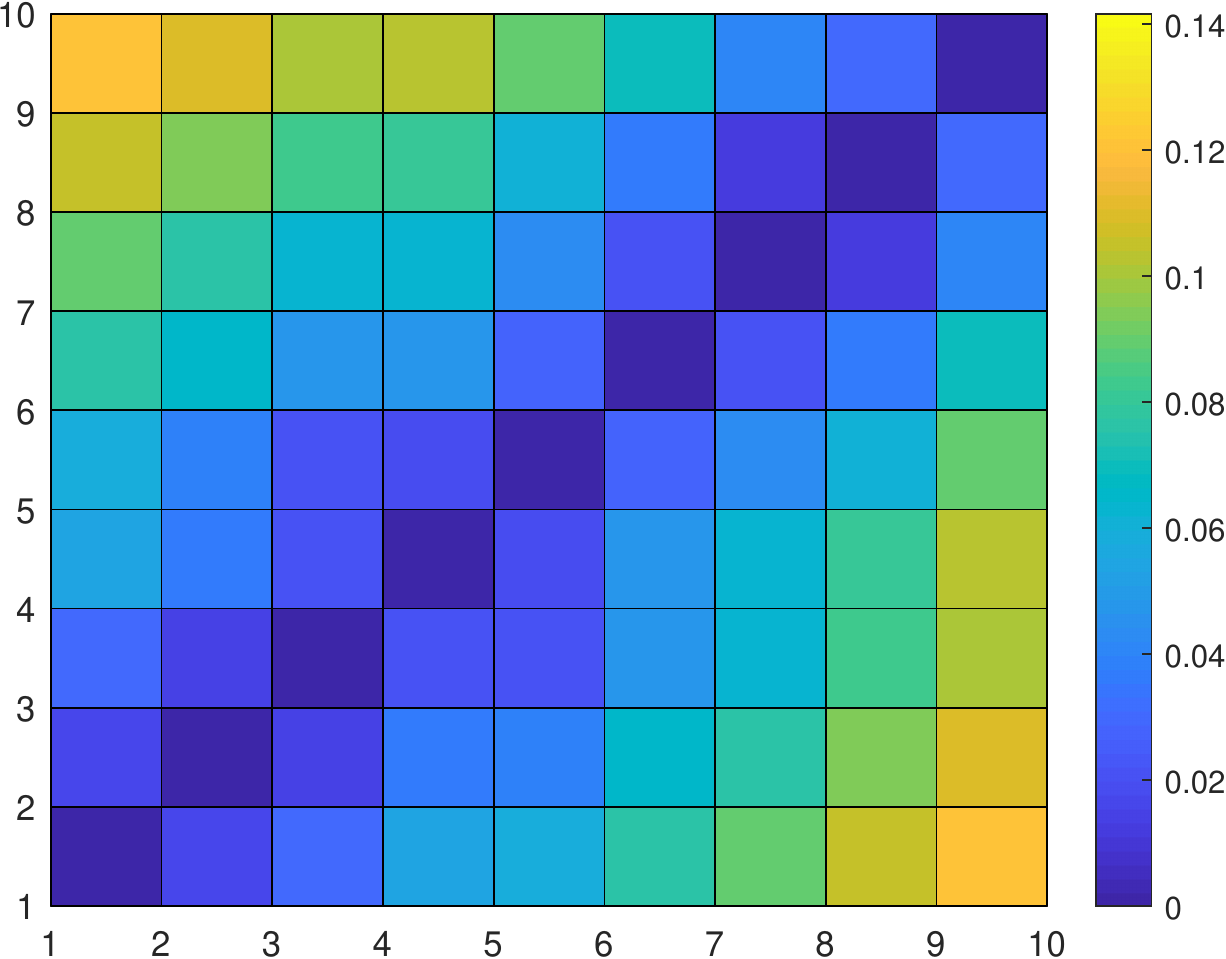}
    \label{fig:temporal correlation d}}
    \subfigure[OB-AD]{
    \includegraphics[width=0.23\linewidth]{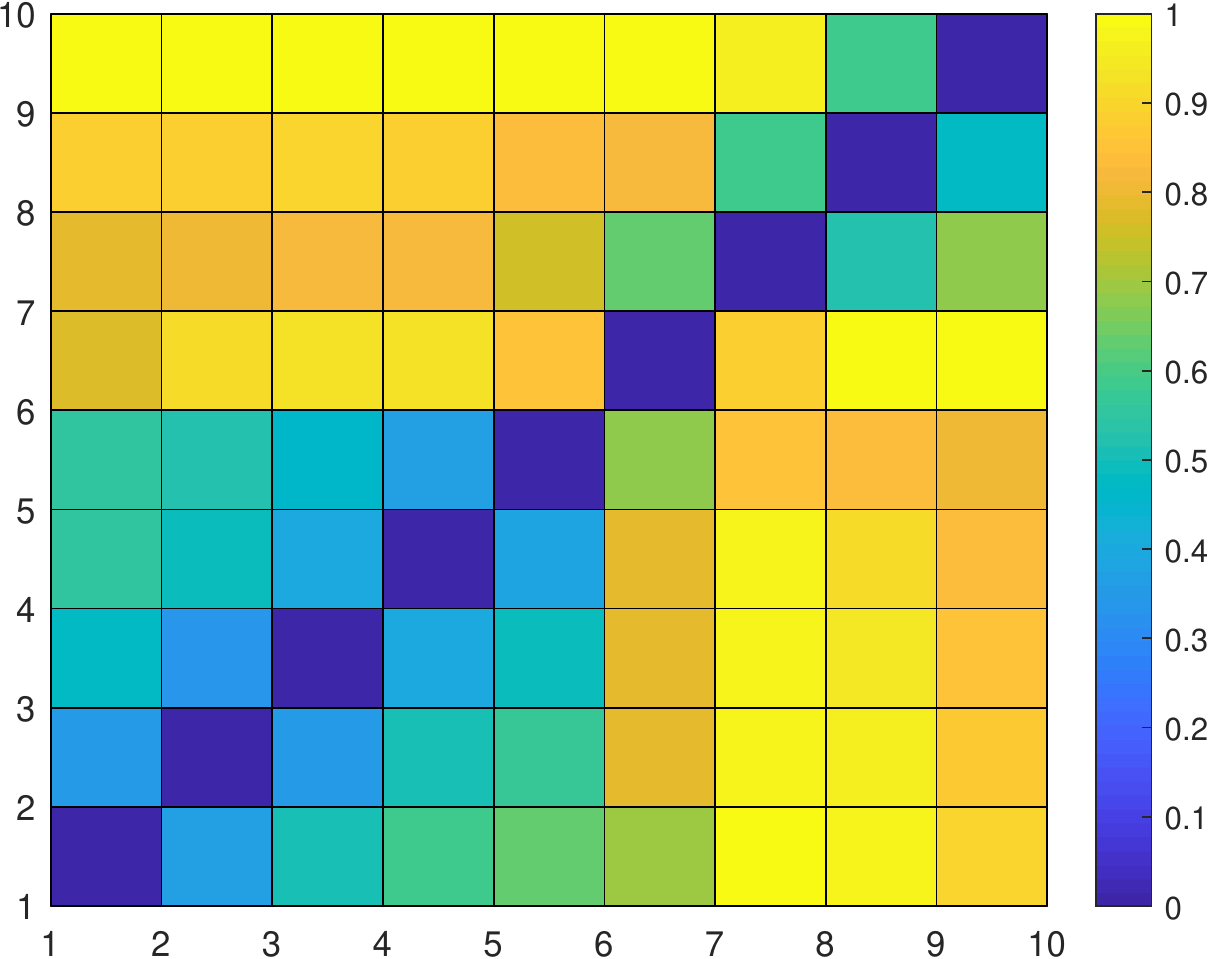}
    \label{fig:temporal correlation e}}
    \subfigure[OB-ED]{
    \includegraphics[width=0.23\linewidth]{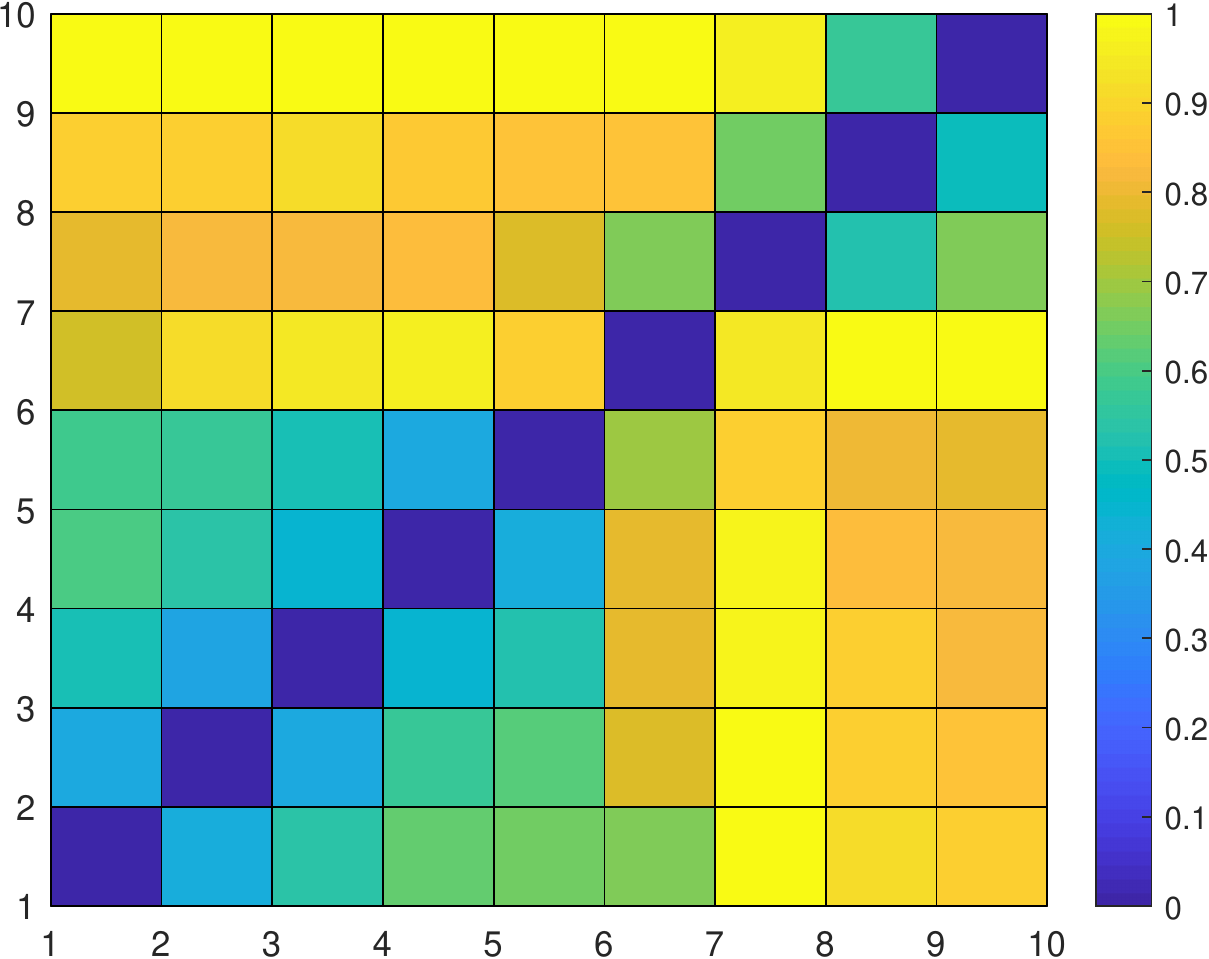}
    \label{fig:temporal correlation f}}
    \subfigure[OB-CDDTW]{
    \includegraphics[width=0.23\linewidth]{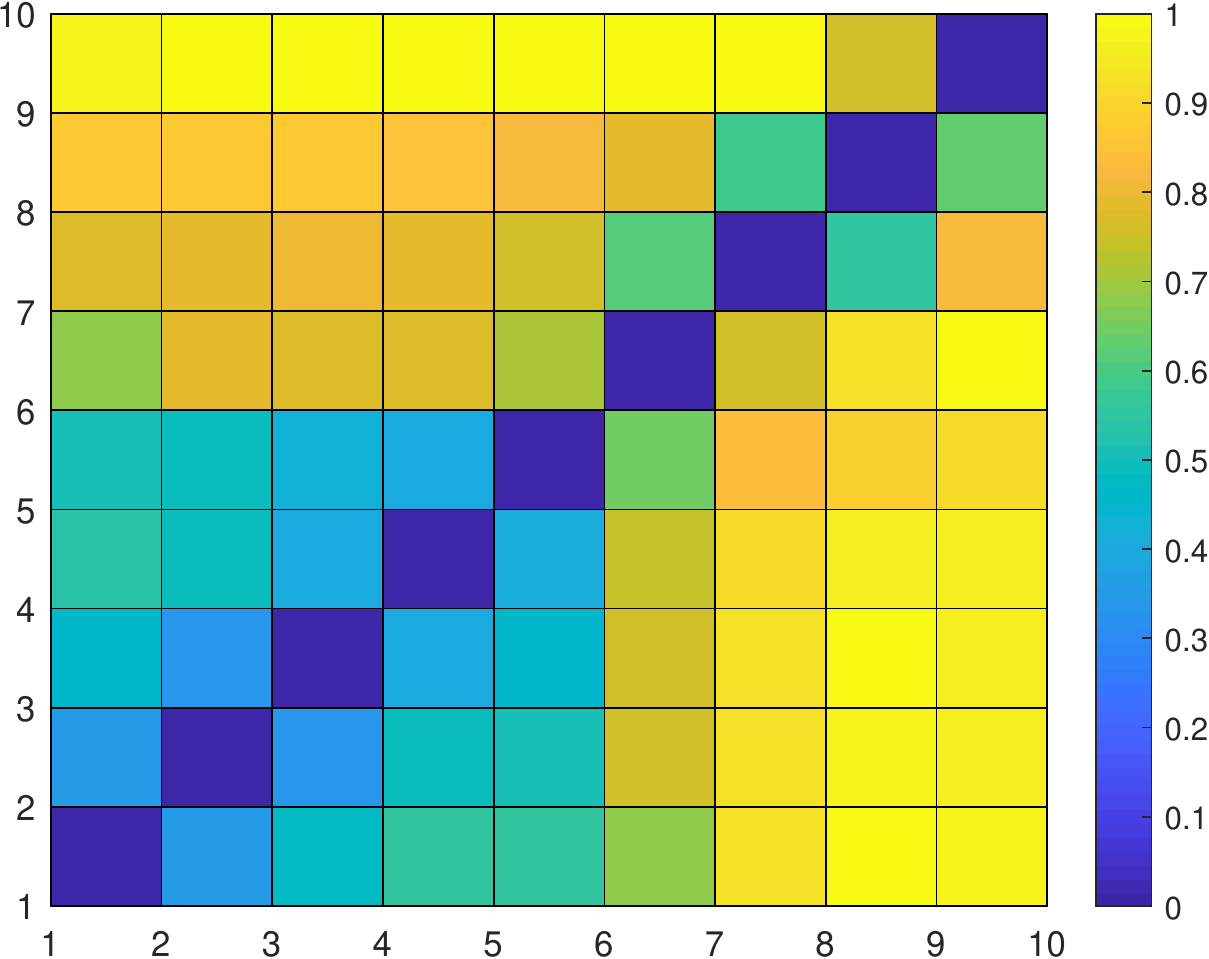}
    \label{fig:temporal correlation g}}
    \subfigure[OB-PD]{
    \includegraphics[width=0.23\linewidth]{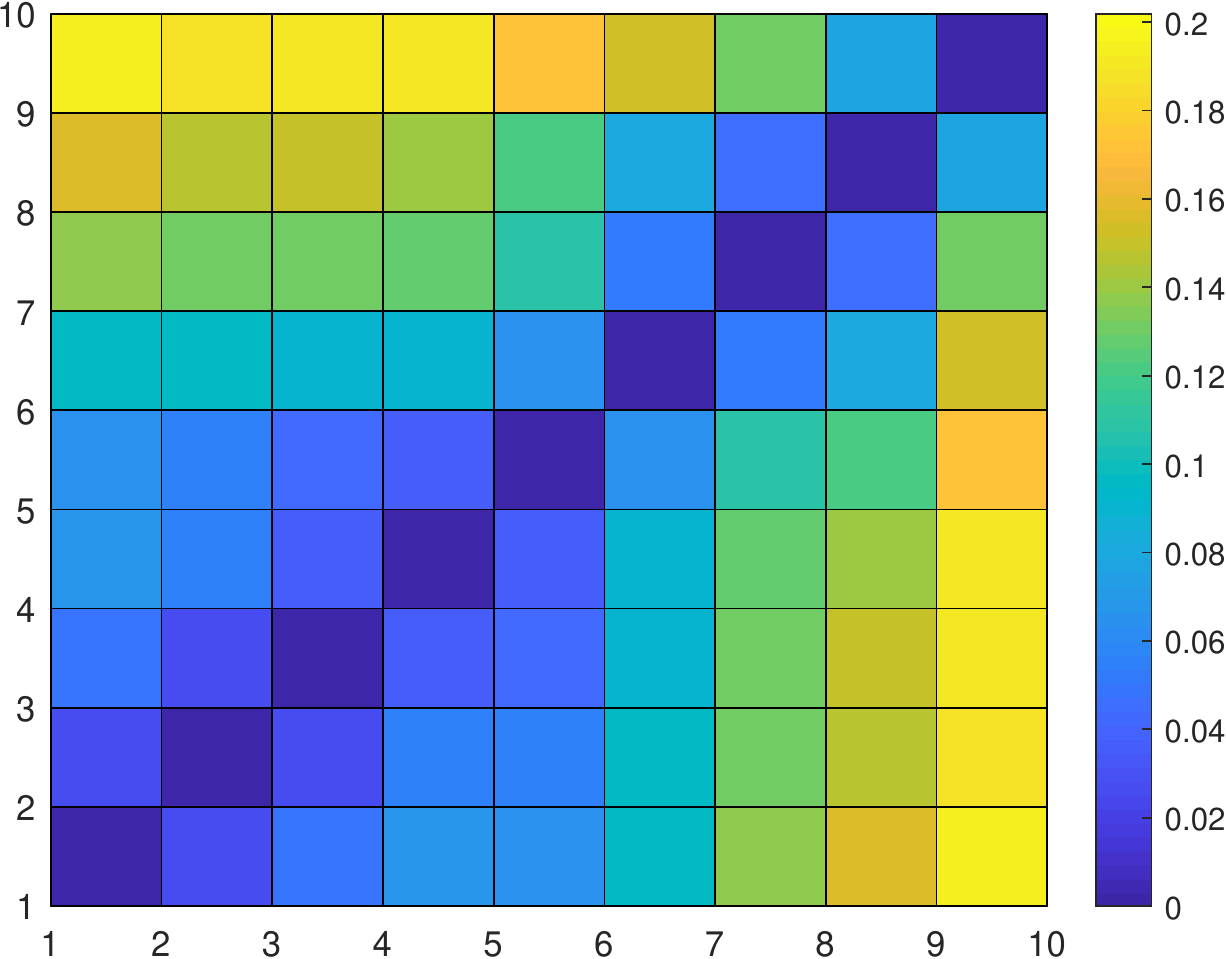}
    \label{fig:temporal correlation h}}
    \caption{Morbidity temporal correlations of COPD and OB (brighter colors indicate bigger difference)}
    \label{fig:temporal correlation}
\end{figure}

To investigate the temporal correlations of morbidity rate for each ward, we calculate the difference of morbidity of one year with the rest of nine years. Then, we apply the same correlation coefficients as above, including AD, ED, PD and CDDTW, to quantify the temporal correlation. The results are shown in Figure~\ref{fig:temporal correlation}. Specifically, Figure~\ref{fig:temporal correlation a} and \ref{fig:temporal correlation e} present the mean value of absolute prevalence difference of the two diseases and the rest of the figures reveal the corresponding correlation distances of every two years for all ten years. The first scale in these figures represents the year of 2009, and the 10th scale represents the year of 2018. The brightness of color reflects the strength of the correlation, the brighter color indicates the weaker temporal correlations.

Figure~\ref{fig:temporal correlation} shows that the brightness of color always increase with the increase of time span regardless of the type of diseases, which empirically verifies that the temporal correlation of morbidity does exit and the prevalence correlation is the strongest in the adjacent years. However, the two figures also present some differences. Figure~\ref{fig:temporal correlation a}, \ref{fig:temporal correlation b}, \ref{fig:temporal correlation c} and \ref{fig:temporal correlation d} show that the brightness of color increases regularly with the increase of time span, but the Figure~\ref{fig:temporal correlation e}, \ref{fig:temporal correlation f}, \ref{fig:temporal correlation g} and \ref{fig:temporal correlation h} show that there are some significant color jumps from one year to next. For instance, in Figure~\ref{fig:temporal correlation g}, the color changes suddenly from green in the cell of row 5 column 2 to yellow in the cell of row 6 column 2, it represents that there is a correlation gap between the 5th year and the 6th year. In conclusion, the temporal correlations are generally exist but there are also some gaps in certain years for certain type of diseases.


\subsection{Summary and Explanation}

The above descriptive analysis demonstrates that the spatiotemporal correlations do exist in a certain range, but the correlations would become weaker and finally disappear with the increase of distance or the increase of time span. 

In this case we would like to further explore the possible reason or explanations behind them. In recent years, researchers in public health domain have been focusing on the heath inequality phenomenon~\cite{Bartley04}, in which they have revealed that there are multiple factors leading to the difference of inter-region health status. They found that the key factors include economic development status, quality of health services, population demographics, and so on~\cite{Fang10}. By reviewing these literature and re-examining the problem in this paper, we believe that the spatiotemporal correlations might partially be explained in the perspectives of these factors. In the spatial view, intuitively, the closer two regions are in geographic, the more similar they are in economic status, health service quality and population demographics, thus they tend to have similar population health status, but if two regions are far enough in geographic, their economic status, health service quality and population demographic would are in totally different situations, so that they tend to have completely different population health status. In the temporal view, the social and economic development (e.g., the increase of people’s income, the improvement of clinical facilities, etc.) for a certain region will be gradual with a certain trend, thus changes of population health status will be gradual with relatively stable trend.

\section{Data Entries Completion for IF-A}
Since the above analysis reveals that the spatiotemporal correlations do exist but are rather complicated, we attempt to investigate if such correlations are strong enough to support the data inference in the IF-A. To achieve this goal, we mathematically formulate the data inference in the IF-A as the problem of missing data entry completion, and then apply a set of mainstream missing data recovery algorithms to verify the completion accuracy. The applied set of algorithms include User-based collaborative filtering (UCF) from spatial perspective, Item-based collaborative filtering (ICF) from temporal perspective, their combination (UCF+ICF), Non-negative Matrix Factorization (NMF) and Higher-Order Tensor Decomposition (HOTD).
\subsection{Spatial View - UCF}
Collaborative Filtering is a data-driven algorithm which has been extensively using in recommendation systems, the general idea behind it is that similar users make similar ratings for similar items~\cite{Su2009}. In this paper we regard a region (here it is a ward) as a user and a time slot (here it is one year) as an item. UCF is a type of CF which applies the similar users' preference to infer others' preferences, so it can be viewed as the utilization of spatial correlation.
\begin{table}[htb]
    \centering
    temporal\\ 
    \begin{tabular}{c|c|c|c|c|c|c|}
        \cline{2-7}
        \multirow{6}{*}{\rotatebox{90}{spatial}}& & $t_{j-2}$ & $t_{j-1}$ & $t_{j}$ & $t_{j+1}$ & $t_{j+2}$\\ 
        \cline{2-7}
         &$s_{i-2}$ & 0.0313 & 0.0314 & 0.031 & 0.032 & 0.0302 \\
         \cline{2-7}
         &$s_{i-1}$ & 0.036 & 0.0356 & 0.03 & 0.03 & 0.0257 \\ 
         \cline{2-7}
         &$s_{i}$ & 0.0298 & 0.0276 & ? & 0.025 & 0.0225 \\ 
         \cline{2-7}
         &$s_{i+1}$ & 0.0321 &  & 0.029 & 0.028 & 0.028 \\ 
         \cline{2-7}
         &$s_{i+2}$ & 0.0296 & 0.0314 & 0.033 & 0.032 &  \\ 
         \cline{2-7}
    \end{tabular}
    \caption{A window of data matrix}
    \label{tab:CFMatrix}
\end{table}

As shown in Table~\ref{tab:CFMatrix}, we construct a window in the data matrix for the missing value \(v_{i,j}\) in the center blank cell with adjacent values \([v_{i,j-\frac{w-1}{2}},v_{i,j+\frac{w-1}{2}}]\), where \(w\) is the window size (in this experiment we test different \(w\) and find \(w=5\) achieve the best performance). Through using the similarity of two regions as a weight we can calculate a weighted average of target entry~\cite{STMVL16}. For example the similarity between \((s_{i},s_{i-1})\) is measured by Equation~\eqref{UCF_similarity}:
\begin{equation}
    sim(s_{i},s_{i-1})={1}/{\sqrt{\frac{\sum_{k=j-\frac{w-1}{2}}^{j+\frac{w-1}{2}} (v_{i,k}-v_{i-1,k})^2}{NT}}}
    \label{UCF_similarity}
\end{equation}
where \(NT\) is the number of timestamps that two regions both have values. The missing value can be inferred according to Equation~\eqref{UCF_weightedaverage}:
\begin{equation}
    \hat{v}^{s}_{i,j}=\frac{\sum_{k=i-\frac{w-1}{2}}^{i+\frac{w+1}{2}} v_{k,j}sim_{k}}{\sum_{k=i-\frac{w-1}{2}}^{i+\frac{w+1}{2}} sim_{k}}
    \label{UCF_weightedaverage}
\end{equation}

\subsection{Temporal View - ICF}
Item-based collaborative filtering (ICF) a form of collaborative filtering for recommendation systems based on the similarity between items calculated using people's ratings of those items. In our formulated problem, we regard regions as users and time-slots as items. Thus, it represents that use the values of adjacent time-slots to infer current value, so it can be viewed as the utilization of temporal correlation. 
Also based on the data matrix constructed in the above UCF model, ICF calculates the similarity between two time-slots according to Equation~\eqref{ICF_similarity} and then uses the similarity as a weight to calculate the weighted average of target entry~\cite{STMVL16} according to Equation~\eqref{ICF_weightedaverage}:
\begin{equation}
    sim(t_{j},t_{j-1})={1}/{\sqrt{\frac{\sum_{k=i-\frac{w-1}{2}}^{i+\frac{w-1}{2}} (v_{k,j}-v_{k,j-1})^2}{NS}}}
    \label{ICF_similarity}
\end{equation}
where \(NS\) is the number of regions that two timestamps both have values.
\begin{equation}
    \hat{v}^{t}_{i,j}=\frac{\sum_{k=j-\frac{w-1}{2}}^{j+\frac{w+1}{2}} v_{i,k}sim_{k}}{\sum_{k=j-\frac{w-1}{2}}^{j+\frac{w+1}{2}} sim_{k}}
    \label{ICF_weightedaverage}
\end{equation}

\subsection{Combination of UCF and ICF (UCF + ICF)}
As UCF mainly utilizes the spatial correlation while ICF mainly utilizes the temporal correlation, we use the combination of UCF and ICF to take advantage of both spatial and temporal correlations. Specifically, it integrates the UCF and ICF to generate a final result according to Equation~\eqref{UCF+ICF}:
\begin{equation}
    \hat{v}^{st} = \lambda_1\hat{v}^t + \lambda_2\hat{v}^s
    \label{UCF+ICF}
\end{equation}

In this paper, we use a set of \(\lambda_1\) and \(\lambda_2\) found by the least square method which can minimise the errors.

\subsection{Non-negative Matrix Factorization (NMF)}
Recently Non-negative Matrix Factorization (NMF)~\cite{Lee1999} is a popular model which aims to find representation of non-negative data automatically and has been applied successfully for dimensional reduction and unsupervised learning.

Given an \(n \times m\) data matrix \(V\) with \(V_{ij}\geq0\) and a pre-specified positive integer \(r<min(n,m)\), NMF finds two non-negative matrices \(W\in R^{n \times r}\) and \(H\in R^{r \times m}\) such that
\begin{equation}
    V \approx WH
    \label{NMF}
\end{equation}

Through finding \(W\) and \(H\) by minimizing the difference between \(V\) and \(WH\) according to Equation~\eqref{NMF_OPTIMIZATION}\cite{Lin07}, we can recover the complete matrix and get the approximation of missing value.
\begin{equation}
    \min_{W,H} f(W,H)\equiv \frac{1}{2}\sum_{i=1}^{n}\sum_{j=1}^{m}\big(V_{ij}-(WH)_{ij}\big)^2
    \label{NMF_OPTIMIZATION}
\end{equation}

In this paper, we divide the original chronic diseases prevalence dataset into multiple data matrices, and each matrix represents one specific type of disease, with the morbidity rate in different regions and for each year. Here, we use NMF to execute data inference of all matrices separately. 

\subsection{Higher-order Tensor Decomposition (HOTD)}
\begin{figure}[htb]
    \centering
    \includegraphics[width=0.5\linewidth]{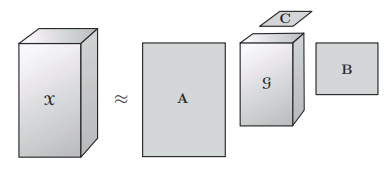}
    \caption{Tucker decomposition}
    \label{fig:Tucker decomposition}
\end{figure}

High-order tensor is defined as \(N\)-way arrays with \(N\geq3\)~\cite{Cichocki2009}, the decomposition of higher-order tensors (HOTD) has extensive applications in signal processing, computer vision, data mining and so forth. Tensor decomposition can also be used for data inference. In this paper, we construct an three-way tensor $\mathcal{X}$ in which the three dimensions represent regions, diseases and times, respectively. It can be factorized into a core tensor $\mathcal{G}$ multiplied by a matrix along each mode \(A\), \(B\) and \(C\)~\cite{Kolda09} according to Equation~\eqref{TuckerDecomposition} as shown in Figure~\ref{fig:Tucker decomposition}:
\begin{equation}
    \mathcal{X} \approx \mathcal{G} \times _1 A \times _2 B \times _3 C
    \label{TuckerDecomposition}
\end{equation}
By minimizing the difference between $\mathcal{X}$ and the multiplication of $\mathcal{G}$, \(A\), \(B\) and \(C\) according to Equation~\eqref{TuckerLossFunction}, we can find $\mathcal{G}$, \(A\), \(B\) and \(C\) and get a approximation of original tensor $\mathcal{X}$:
\begin{equation}
    \begin{split}
    \min_{\mathcal{G},A,B,C} \mathcal{L}(\mathcal{G},A,B,C)&=\frac{1}{2}\Vert\mathcal{X}-\mathcal{G}\times_A A \times_B B \times_C C\Vert_2 \\ & + \frac{\lambda}{2}(\Vert\mathcal{G}\Vert_2 + \Vert A \Vert_2 + \Vert B \Vert_2 + \Vert C \Vert_2)
    \end{split}
    \label{TuckerLossFunction}
\end{equation}
where \(\Vert\cdot\Vert_2\) represents L2-norm and the second item on the right side is the penalty used to prevent over-fitting~\cite{zheng2014diagnosing}.

\subsection{Inference Performance Comparison}
We execute the aforementioned five algorithms to do data inference separately and compare their differences in inference errors. Specifically, we use two metrics: Root Mean Square Error (RMSE) and Mean Absolute Error (MAE) 
\begin{equation}
    RMSE=\sqrt{\frac{\sum_i{(y_i-\hat{y_i})^2}}{n}}
\end{equation}
\begin{equation}
    MAE=\frac{\sum_i\vert{y_i-\hat{y_i}}\vert}{n}
\end{equation}
where \(\hat{y_i}\) is an inference, \(y_i\) is the gound truth, and \(n\) is the number of inferred missing value.

\begin{figure}[htb]
    \centering
    \subfigure[RMSE]{
    \includegraphics[width=0.4\linewidth]{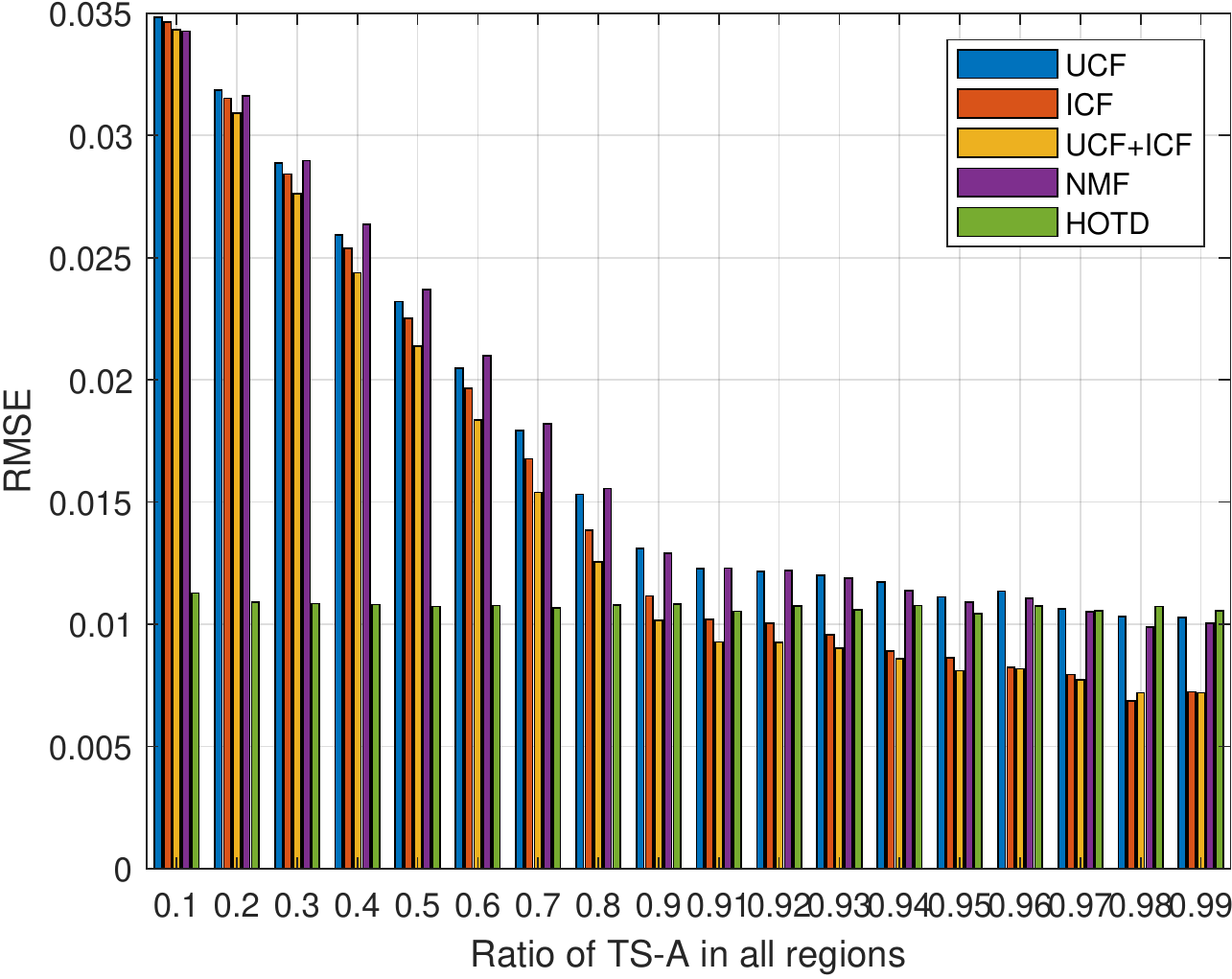}
    \label{fig:RecoveryRMSE}}
    \subfigure[MAE]{
    \includegraphics[width=0.4\linewidth]{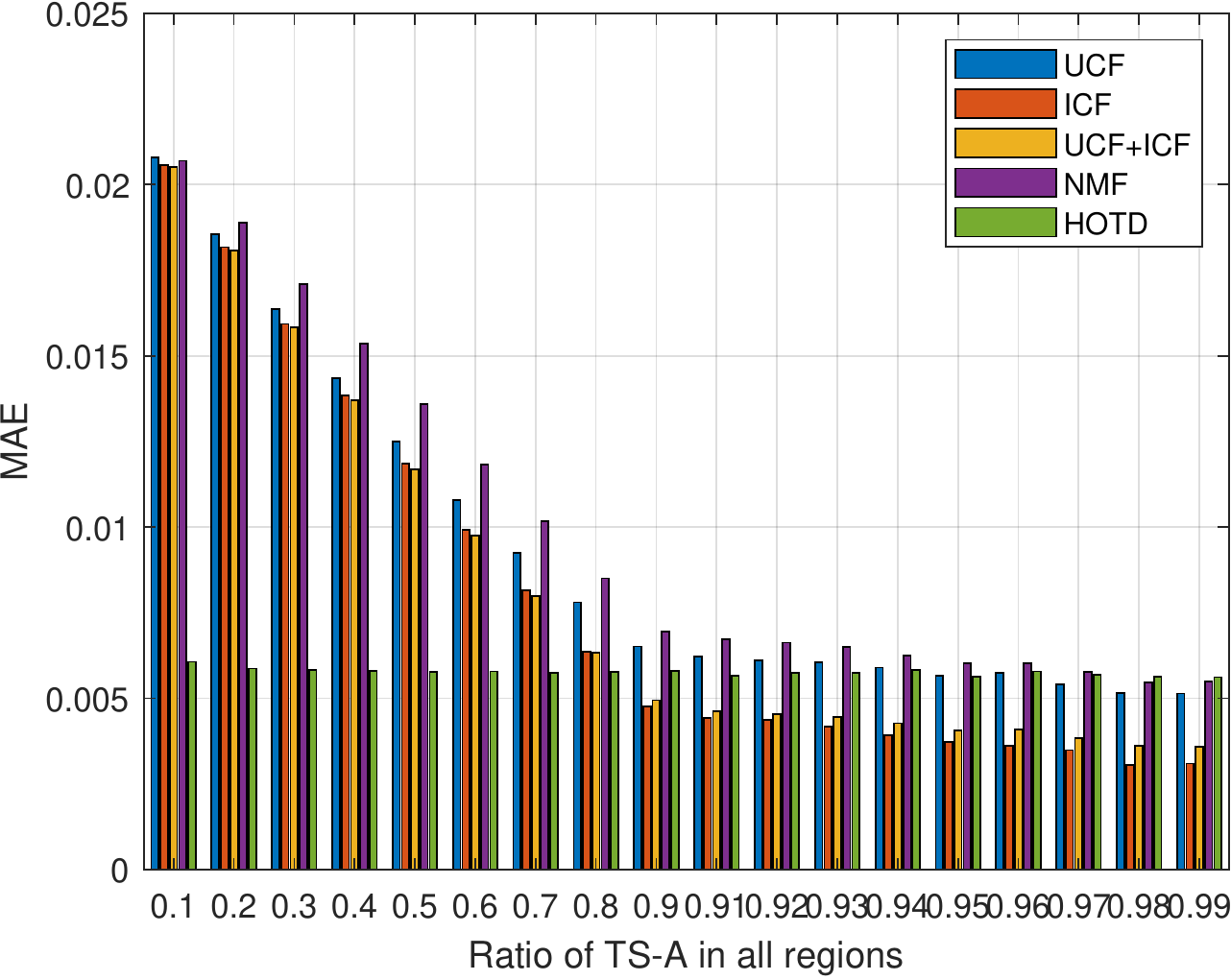}
    \label{fig:RecoveryMAE}}
    \caption{Inference quality comparison with various proportion of TS-A}
    \label{fig:Recovery}
\end{figure}

We make comparisons by varying the proportion of known data (i.e.,ration of TS-A in all regions) from 0.1 to 0.9. Specifically, for each year, we randomly select a proportion of regions, assuming that morbidity rates of all diseases in these regions are available and we infer the missing value of others. Here, in order to keep consistent with the idea of CPH, we only use the data in previous years to do the data inference. For example, to infer the missing entries in the year of 2015, the algorithms are only allowed to use the data from 2009 to 2014. The inferring quality is shown in Figure~\ref{fig:Recovery}.

By observing the results shown in Figure~\ref{fig:Recovery}, we can see that:
\begin{itemize}[leftmargin=*]
    \item All algorithms perform better gradually with the increase of TS-A proportion as more available data facilitates the improvement of inferring accuracy, which is consistent with our intuitions;
    \item The combination of UCF and ICF performs better than UCF and ICF respectively, because integrating spatial and temporal correlation is better than applying any single of them;
    \item HOTD outperforms the other algorithms significantly when the proportion of TS-A is less than 90\%. However, when the proportion exceeds 90$\%$, it loses its advantage compared to some others. Intuitively, HOTD uses global information to infer missing data, which implicitly utilizes the correlations among different types of diseases. On the contrary, other algorithms do not consider these inter-disease correlations. Thus, HOTD demonstrates its superiority over others under most of the data availability settings. However, when most of the data become available, using local information is informative enough for other algorithms to achieve good inference results. In these settings, introducing the inter-disease correlations may have side effects as it brings extra noises.
\end{itemize}

As the above results show the data inference accuracy for different data entry completion algorithms when varying the data availability, we need to go back to our original goal of investigating if such performance is satisfactory for real-world PHM tasks. To accomplish this goal, we first need to know what is the maximum error that healthcare administrators can accept. To our best of knowledge, there are no literatures explicitly defining such thresholds. Thus, we turn to several domain experts (including NHS staffs and professors in public health) for help. Through the communications, we know from the domain experts that this threshold may vary among diseases and countries, but the average value is approximately 0.01 in terms of MAE according to their experiences. With such value in mind, we go back to Figure~\ref{fig:RecoveryMAE}, from which we can see that the HOTD can satisfy such requirement even by selecting only 10\% regions as TS-A. It indicates that the idea of CPH does have potential impact in PHM, which can save at least 90\% budget for data collection. Considering that we can select more informative regions and design more intelligent data entry recovery algorithms, such impact would become more significant in the future. 

\section{Optimized Selection of TS-A}
As the missing values can be inferred accurately from a relatively small subset of TS-A, we attempt to select the more informative regions as TS-A and compare it with random selection method. This problem is similar to the adaptive sampling in sensor networks, which may have two modes, i.e., the online mode and the offline mode. The online mode~\cite{Leye15} selects one location to collect sensing data, and selects the next location after the data of previously selected one returns data. Alternatively, the offline mode~\cite{Guestrin05} selects all locations all at once and then collects data of all locations in a parallel way. In our CPH framework, collecting the data (e.g., morbidity rate) in each ward is very time consuming, so we assume that our CPH follows the offline mode, which utilizes the historical data to select a fixed number wards  (predefined by the budget of the PHM task) as TS-A. 

Therefore, the problem is mathematically formulated as: given the historical morbidity rate data of previous years, we aim to select a predefined proportion of wards  as TS-A with the goal of minimizing the data entry completion errors for IF-A in the current year. Specifically, we adopt the following two algorithms. 


The  first method is called Regions with Maximum Dispersion Coefficient (RMDC), and it selects the predefined proportion of wards with the maximum Dispersion Coefficient (DC) which is defined in equation\eqref{DispersionCoefficietEquation}. DC is originally used as a metric in the discipline of environment engineering, which characterizes the amount of pulse broadening by material dispersion per unit length of fiber and per unit of spectral width. Here, higher DC means that a ward is more informative in terms of data recovery for other wards. 
\begin{equation}
   DC_i = \frac{\sqrt{\sum_{k=i-m}^{i+m}{(V_k-\Bar{V})^2}}}{\Bar{V}}
    \label{DispersionCoefficietEquation}
\end{equation}
where \(m=5\), $DC_i$ represents the dispersion coefficient of the $i$th region, $V_k$ represents the value of  the $k$th adjacent region, and $\Bar{V}$ represent the mean value of all adjacent regions.

The second method we used to select the regions is called Query-by-Committee-based algorithm (QCB). Committee here refers to a set of various missing entry recovery algorithms described in Section 4. For previous years, we cover the value of each ward, and use data in other wards to do data inference by adopting every algorithm in the committee to infer the missing value. Finally, we select wards with the largest variance among the deduced values of different algorithms as TS-A. 

\begin{figure}[htb]
    \centering
    \subfigure[UCF-RMSE]{
    \includegraphics[width=0.23\linewidth]{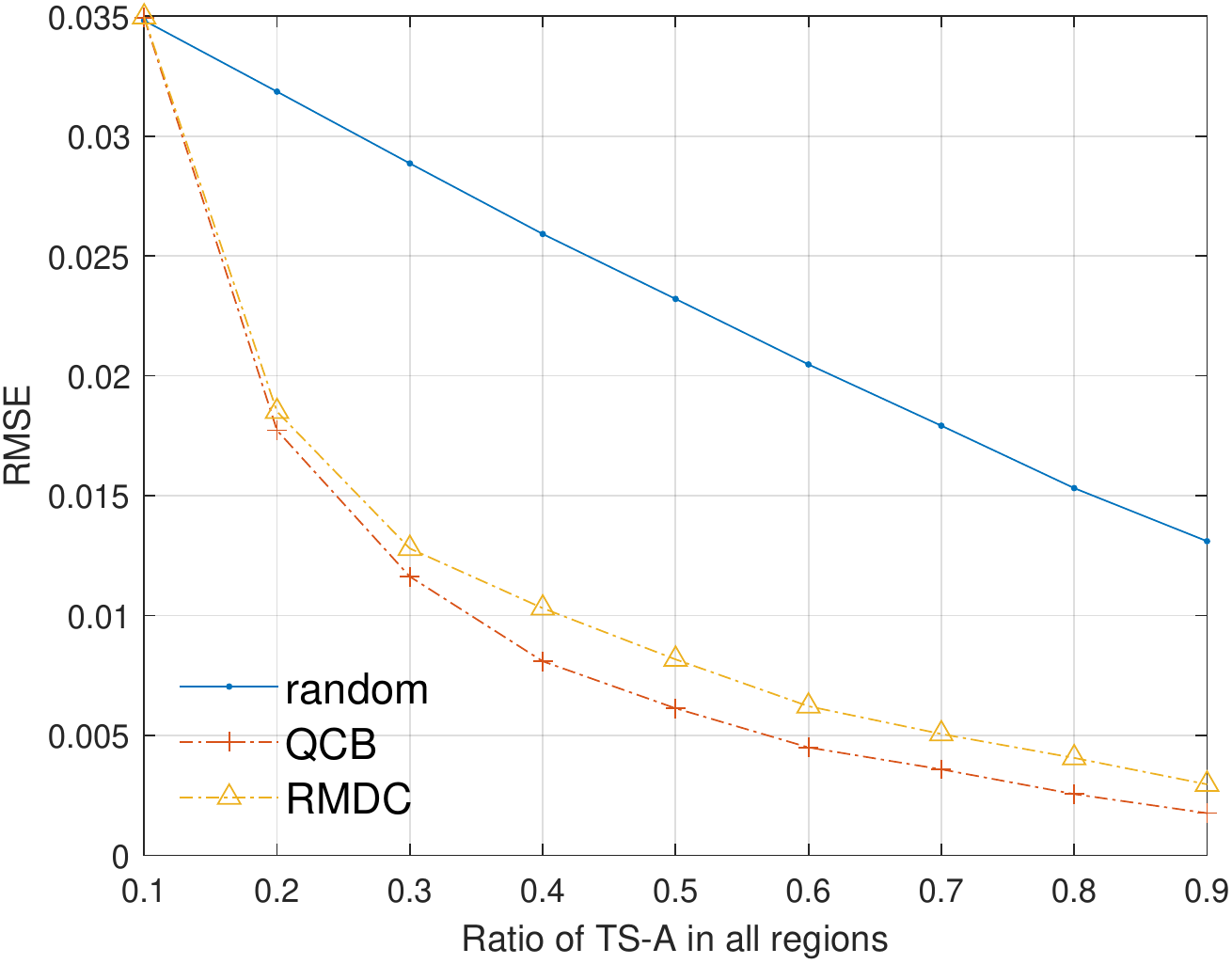}
    \label{fig:PerformanceComparison_a}}
    \subfigure[UCF-MAE]{
    \includegraphics[width=0.23\linewidth]{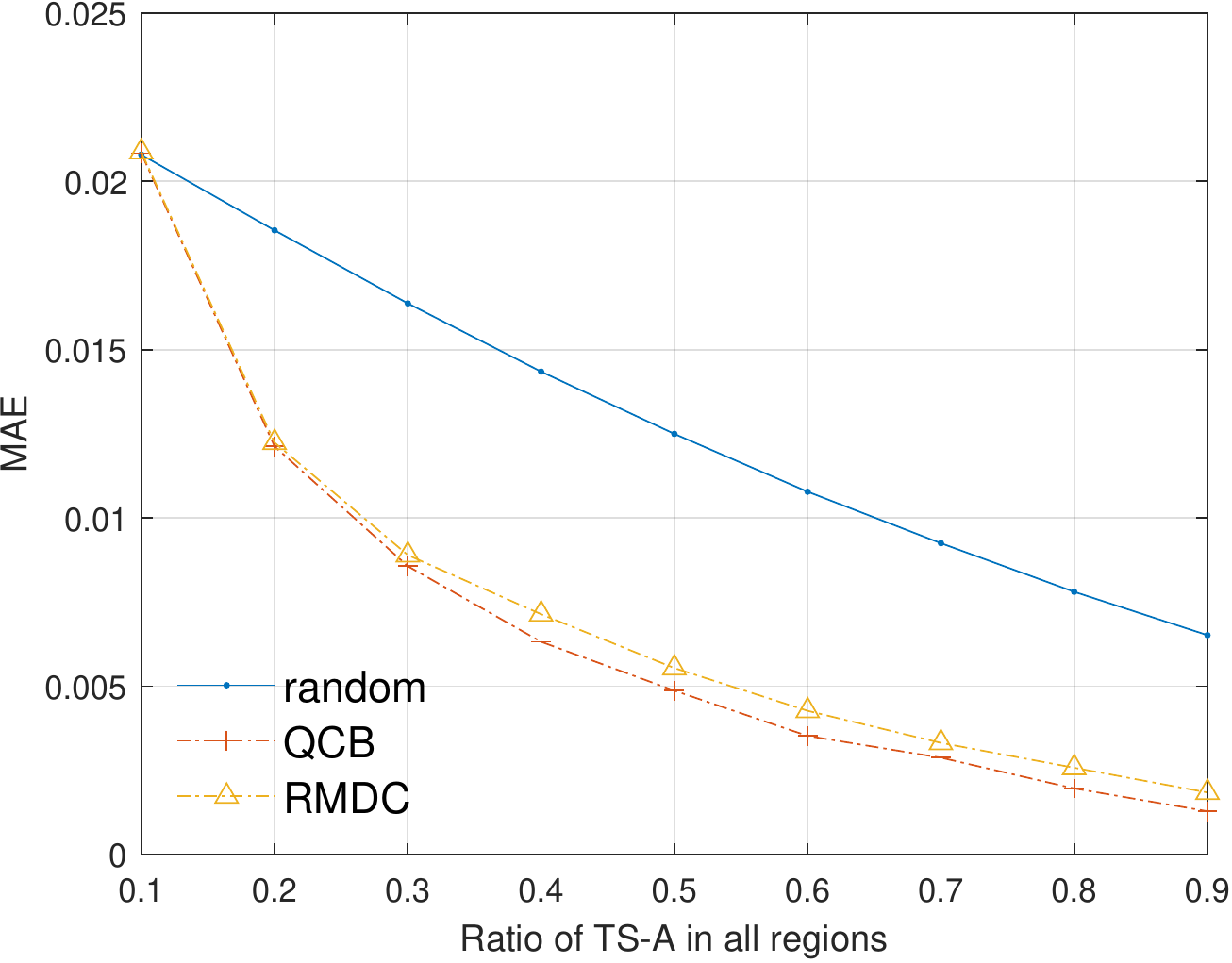}
    \label{fig:PerformanceComparison_b}}
    \subfigure[ICF-RMSE]{
    \includegraphics[width=0.23\linewidth]{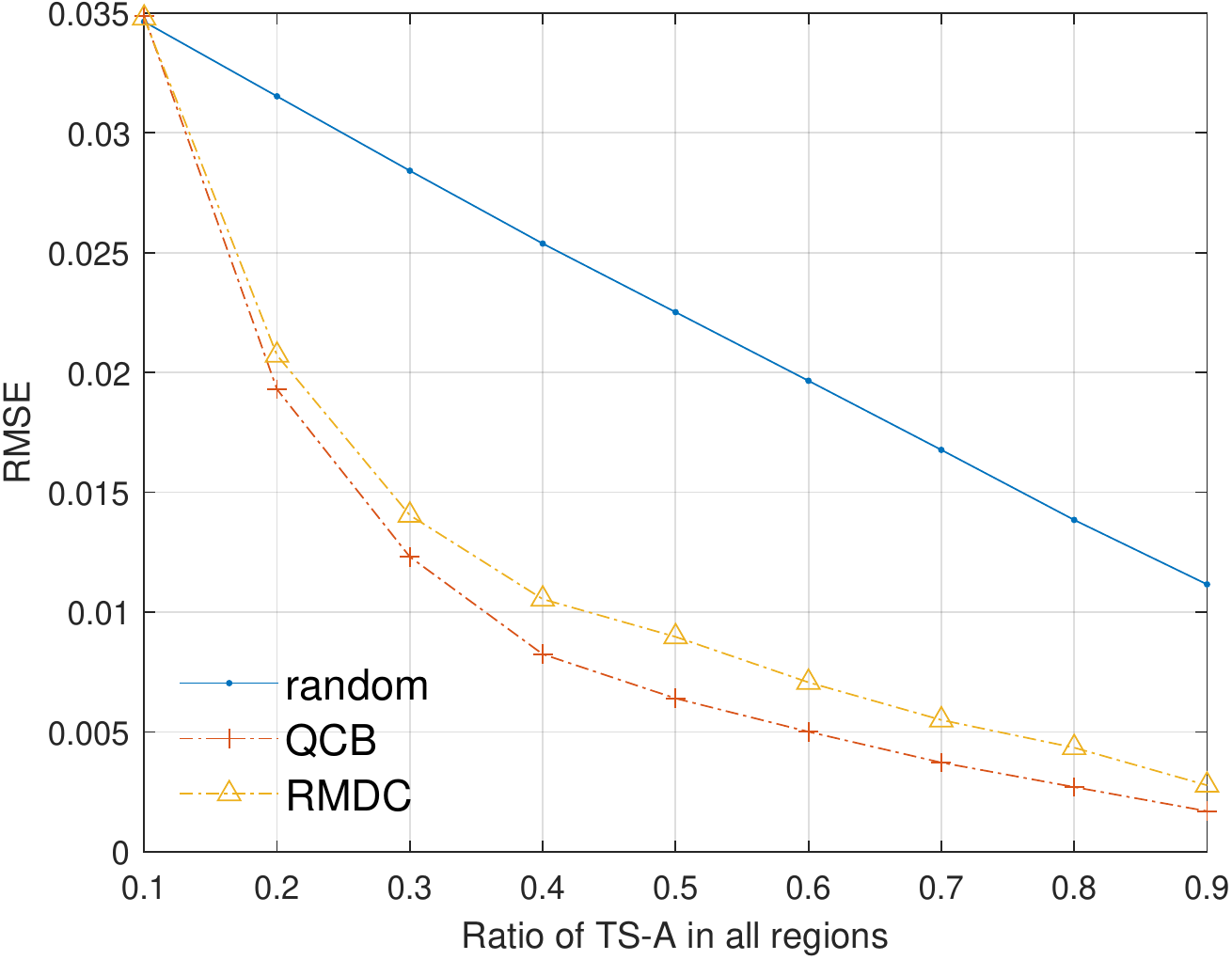}
    \label{fig:PerformanceComparison_c}}
    \subfigure[ICF-MAE]{
    \includegraphics[width=0.23\linewidth]{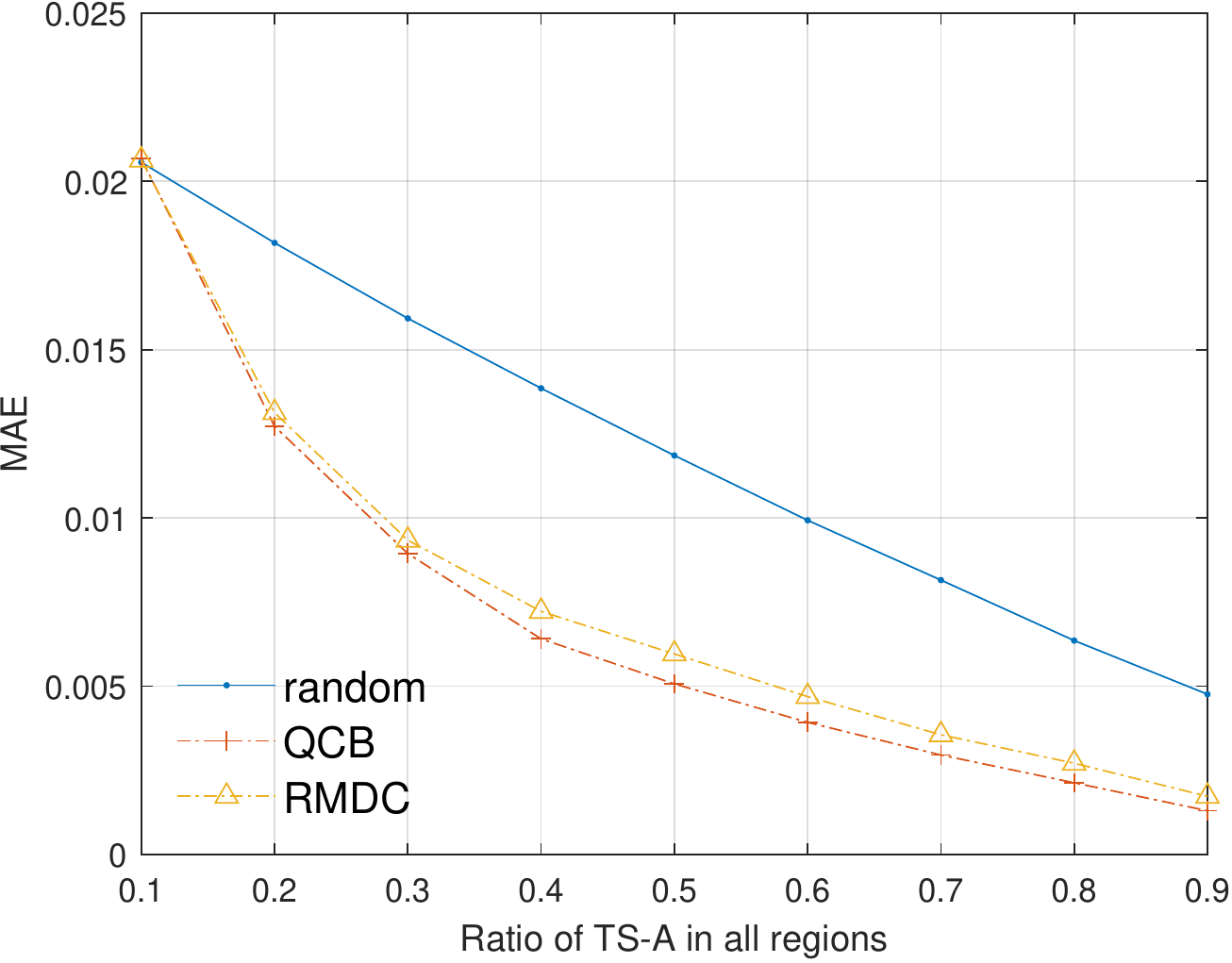}
    \label{fig:PerformanceComparison_d}}
    \subfigure[UCF+ICF-RMSE]{
    \includegraphics[width=0.23\linewidth]{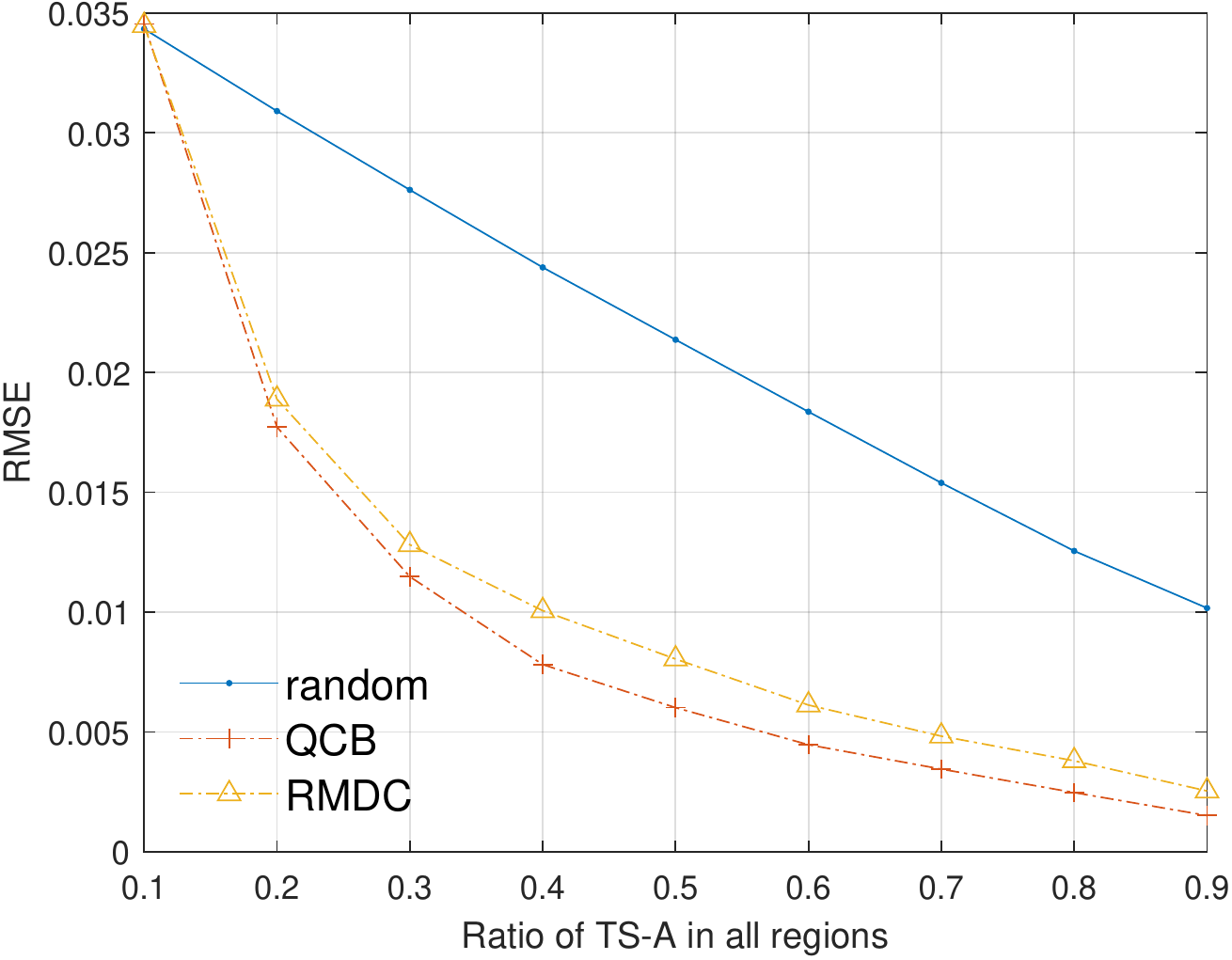}
    \label{fig:PerformanceComparison_e}}
    \subfigure[UCF+ICF-MAE]{
    \includegraphics[width=0.23\linewidth]{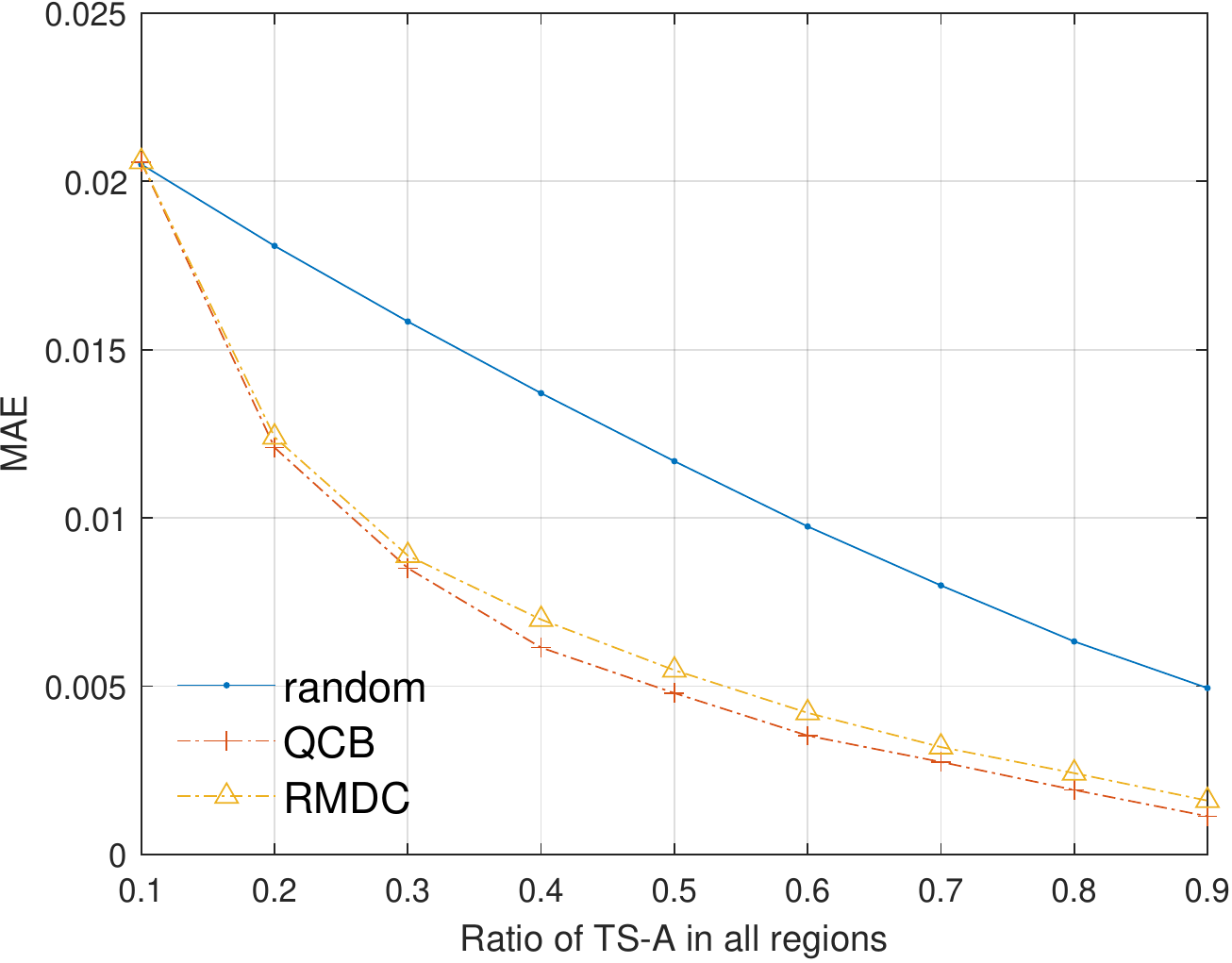}
    \label{fig:PerformanceComparison_f}}
    \subfigure[NMF-RMSE]{
    \includegraphics[width=0.23\linewidth]{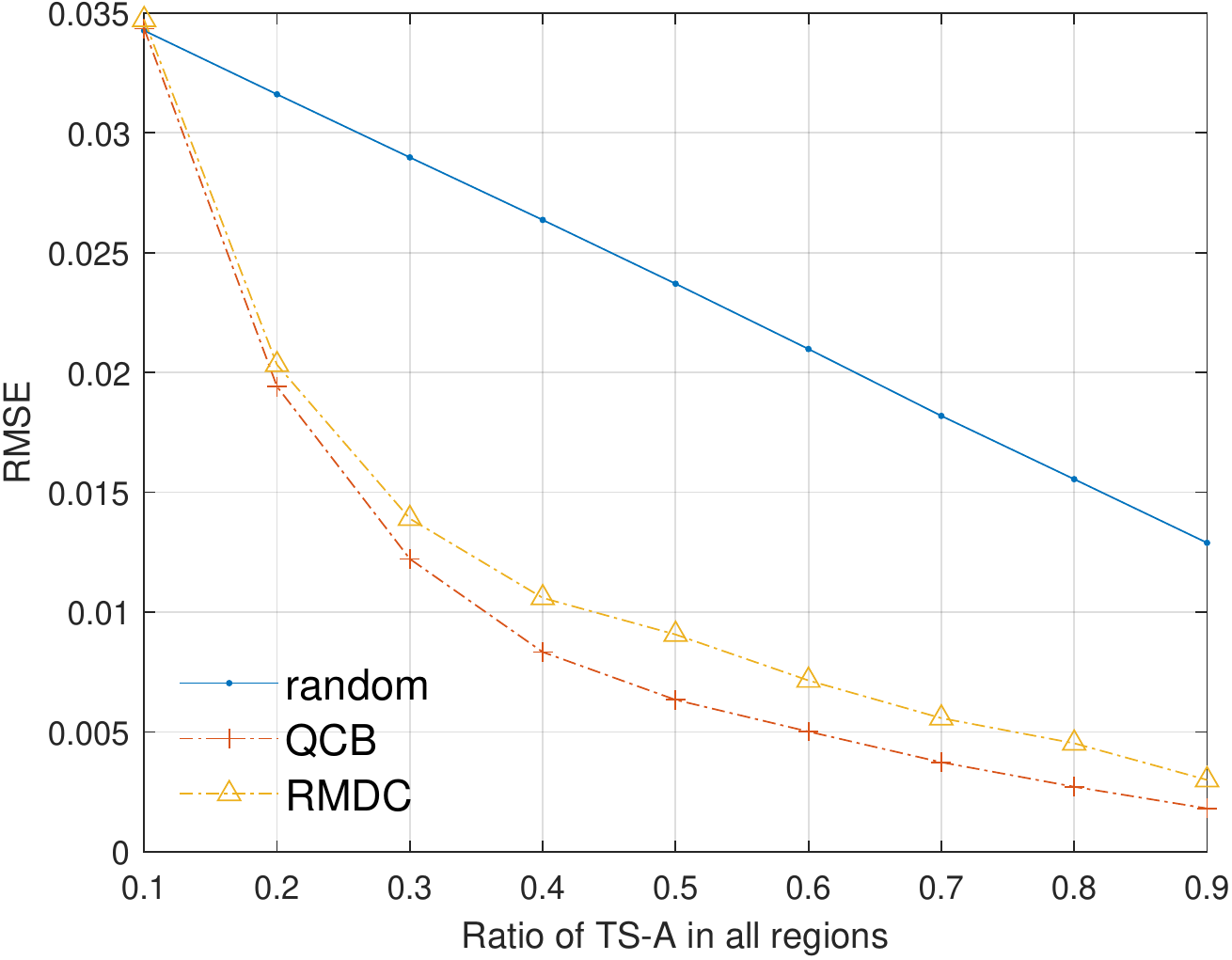}
    \label{fig:PerformanceComparison_h}}
    \subfigure[NMF-MAE]{
    \includegraphics[width=0.23\linewidth]{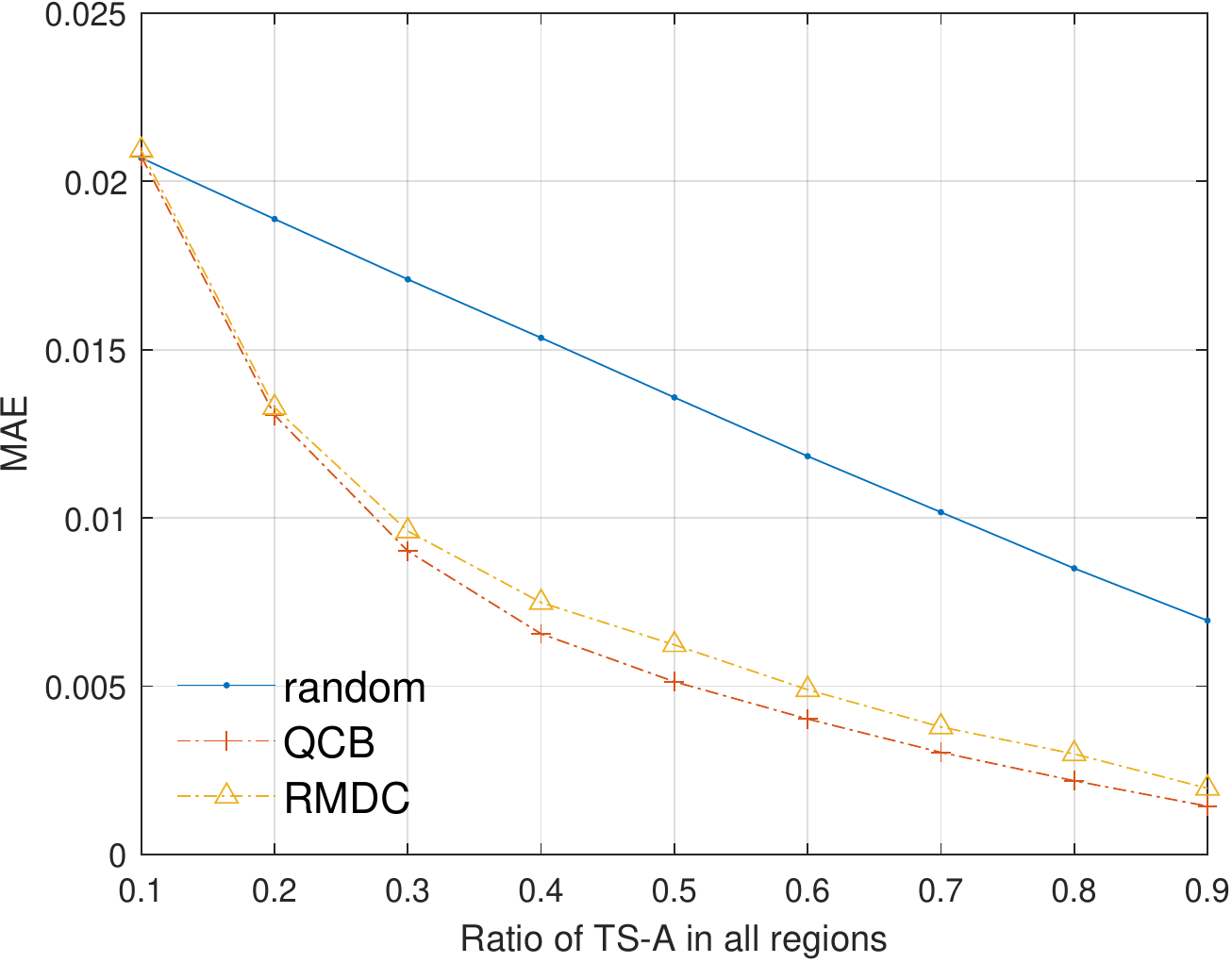}
    \label{fig:PerformanceComparison_i}}
    \subfigure[HOTD-RMSE]{
    \includegraphics[width=0.23\linewidth]{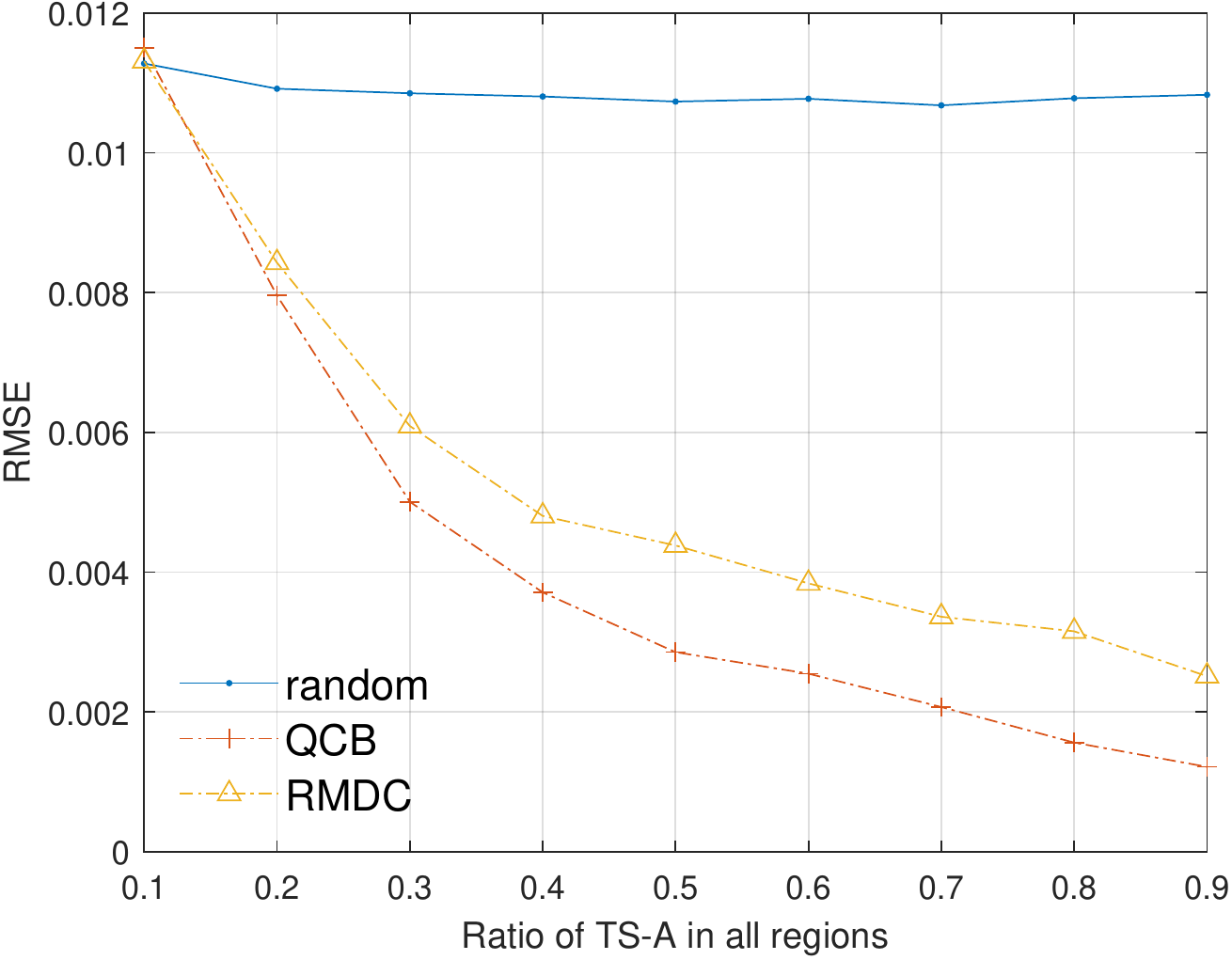}
    \label{fig:PerformanceComparison_j}}
    \subfigure[HOTD-MAE]{
    \includegraphics[width=0.23\linewidth]{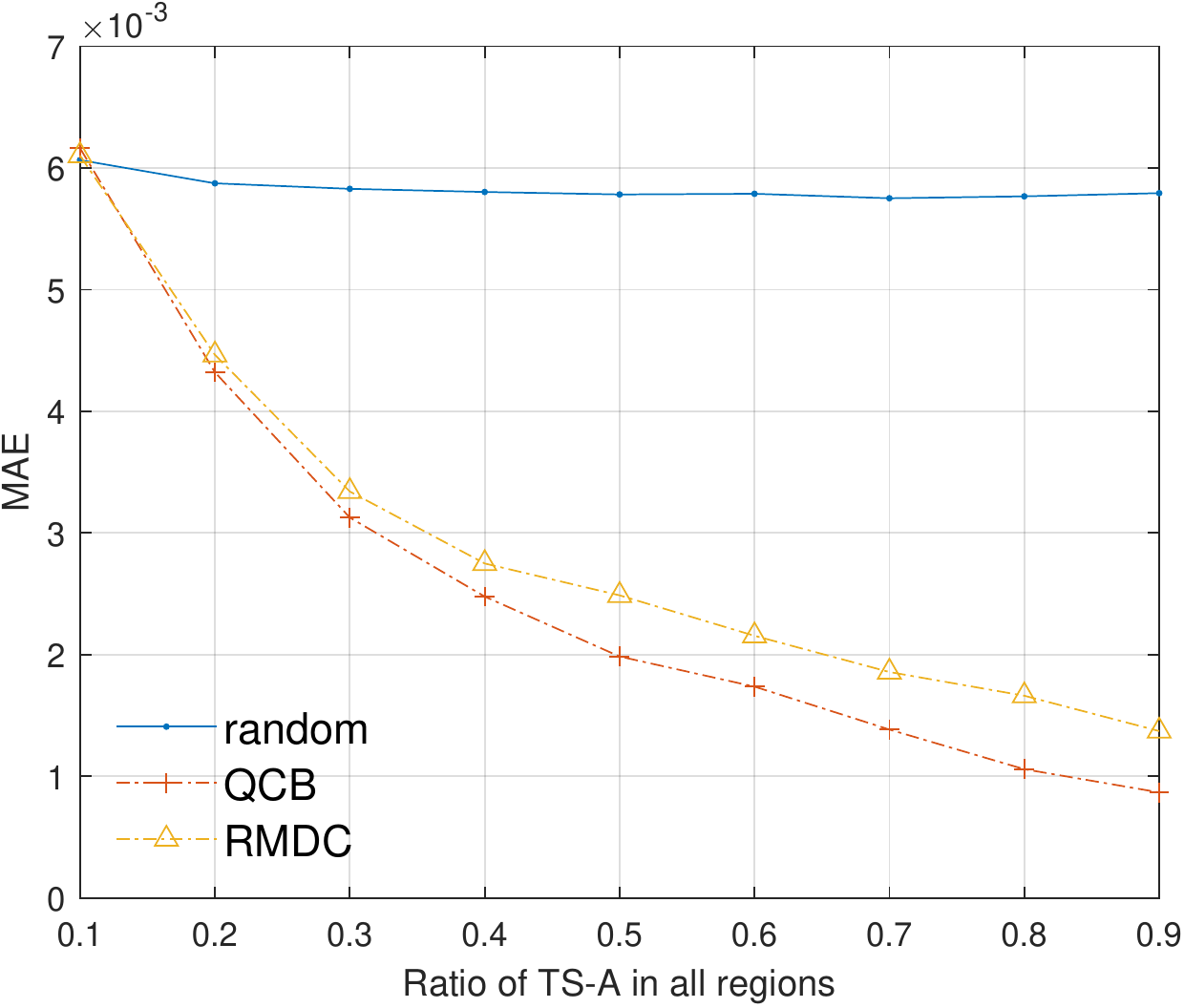}}
    \caption{Performance comparison of all algorithms with the increase of TS-A ratio}
    \label{fig:PerformanceComparison}
\end{figure}

Figure~\ref{fig:PerformanceComparison} shows the comparison of these two methods over random selection approach when varying the proportion of TS-A from 0.1 to 0.9 and changing data entry completion methods.  From figure~\ref{fig:PerformanceComparison}, we can see that: (1) all approaches generally achieves better results (lower errors) as the number of selected TS-A increases. (2) QCB and RMDC outperform the random selection significantly.  (3) We find that QCB consistently outperforms RMDC in all settings, and the advantage becomes the most significant when fixing the HOTD as the data entry completion method. In conclusion, the results suggest that the optimized selection of TS-A is definitely effective and it can significantly improve the implementation of CPH.


\begin{figure}[htb]
    \centering
    \subfigure[QCB-RMSE]{
    \includegraphics[width=0.47\linewidth]{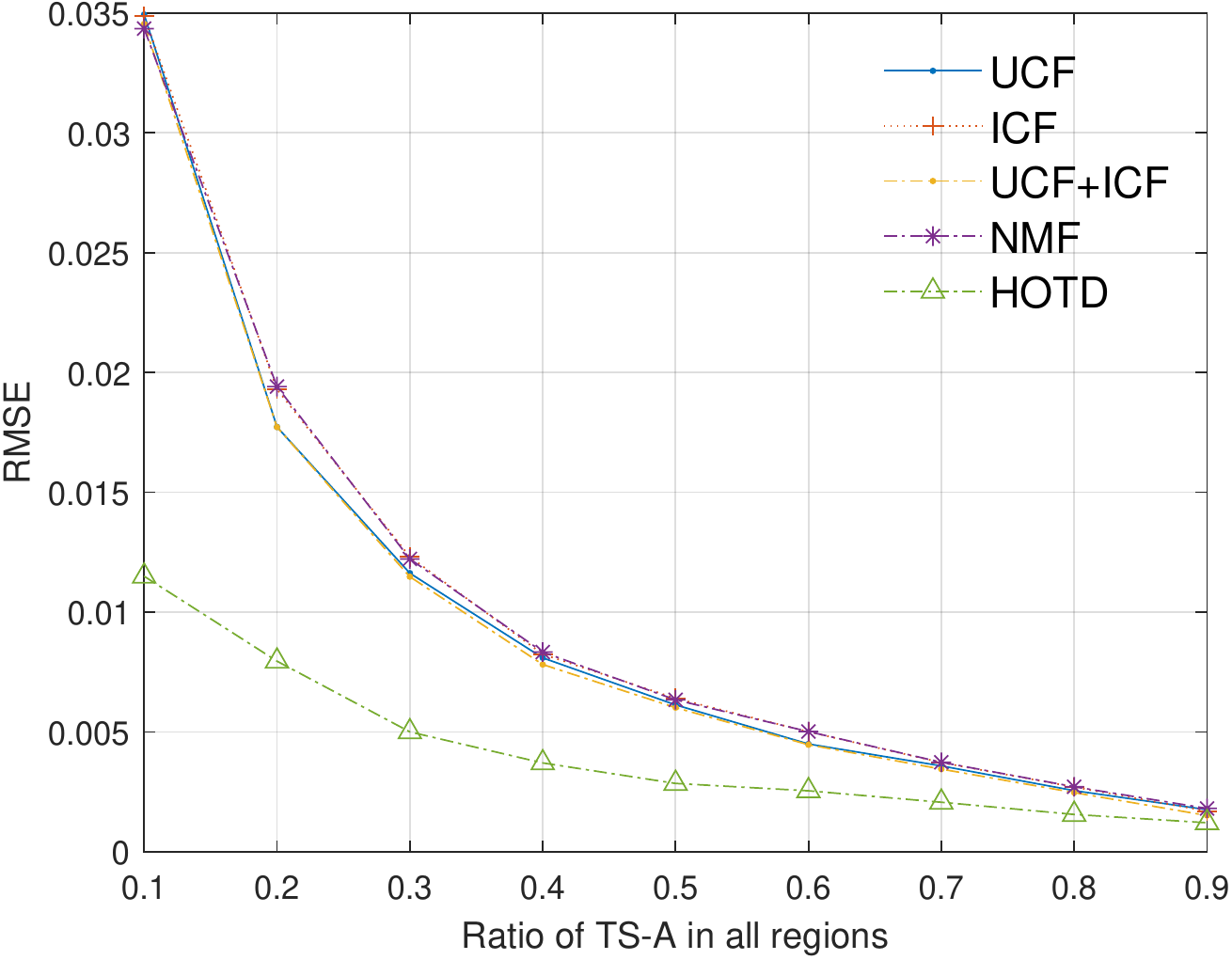}}
    \subfigure[QCB-MAE]{
    \includegraphics[width=0.47\linewidth]{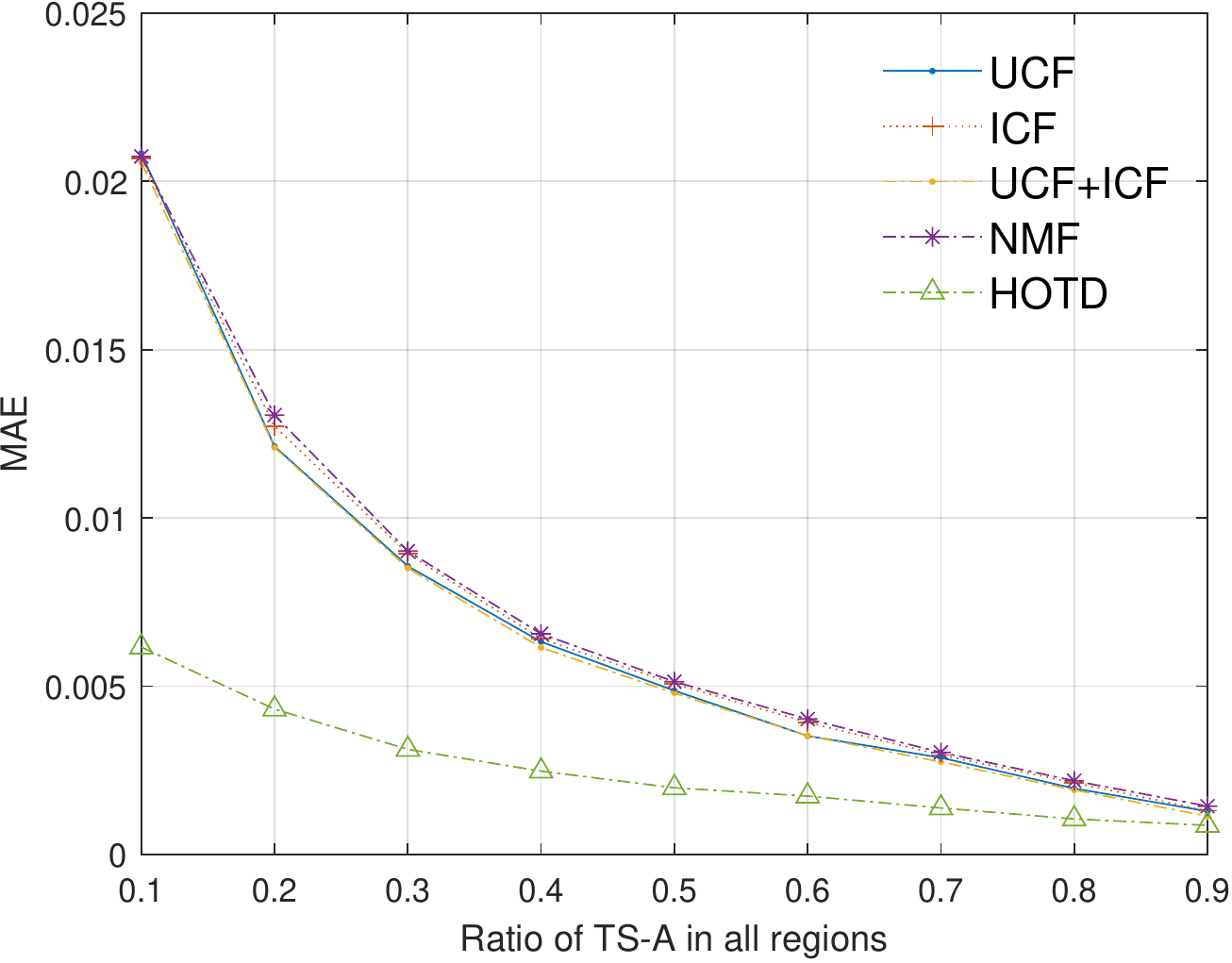}}
    \subfigure[RMDC-RMSE]{
    \includegraphics[width=0.47\linewidth]{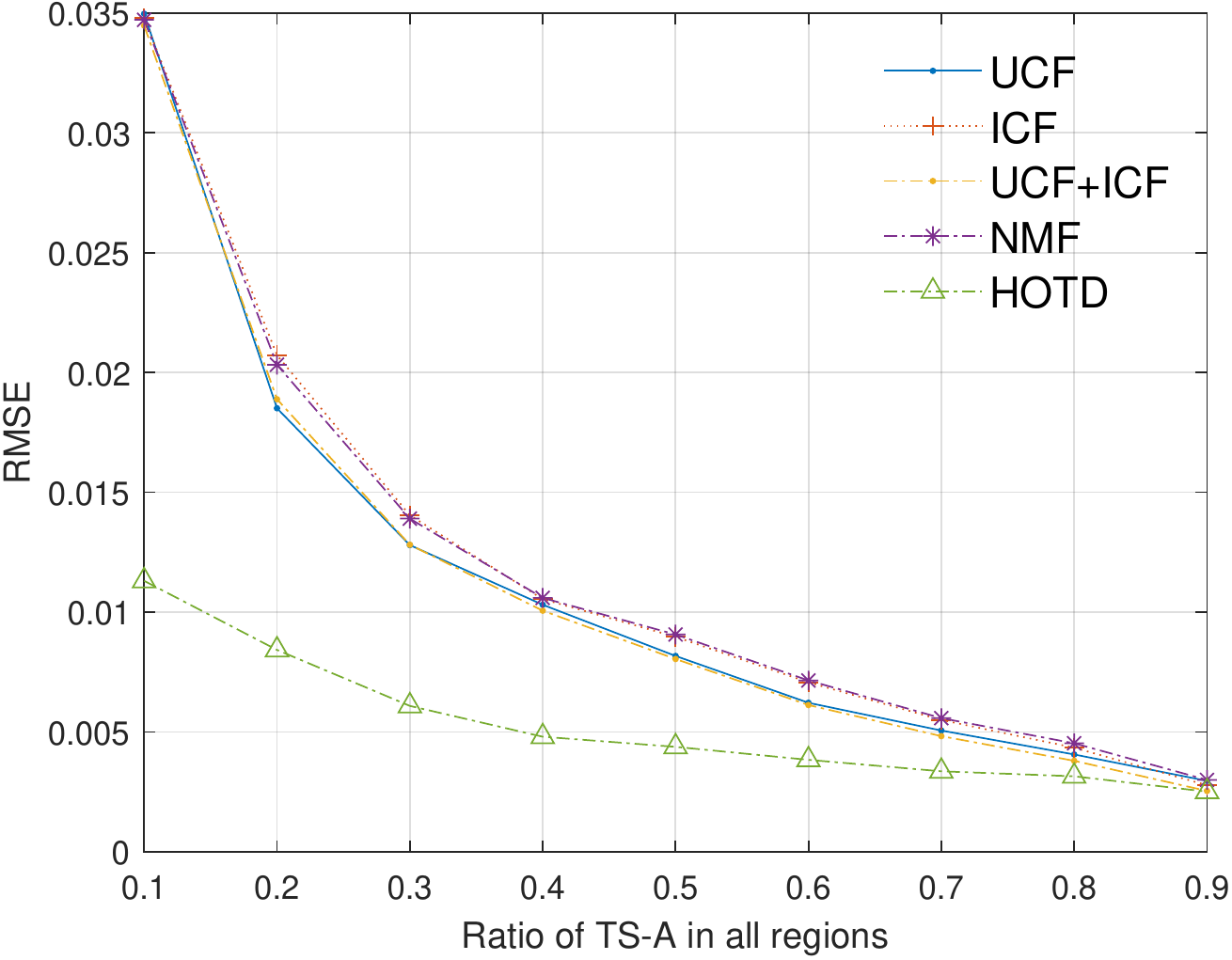}}
    \subfigure[RMDC-MAE]{
    \includegraphics[width=0.47\linewidth]{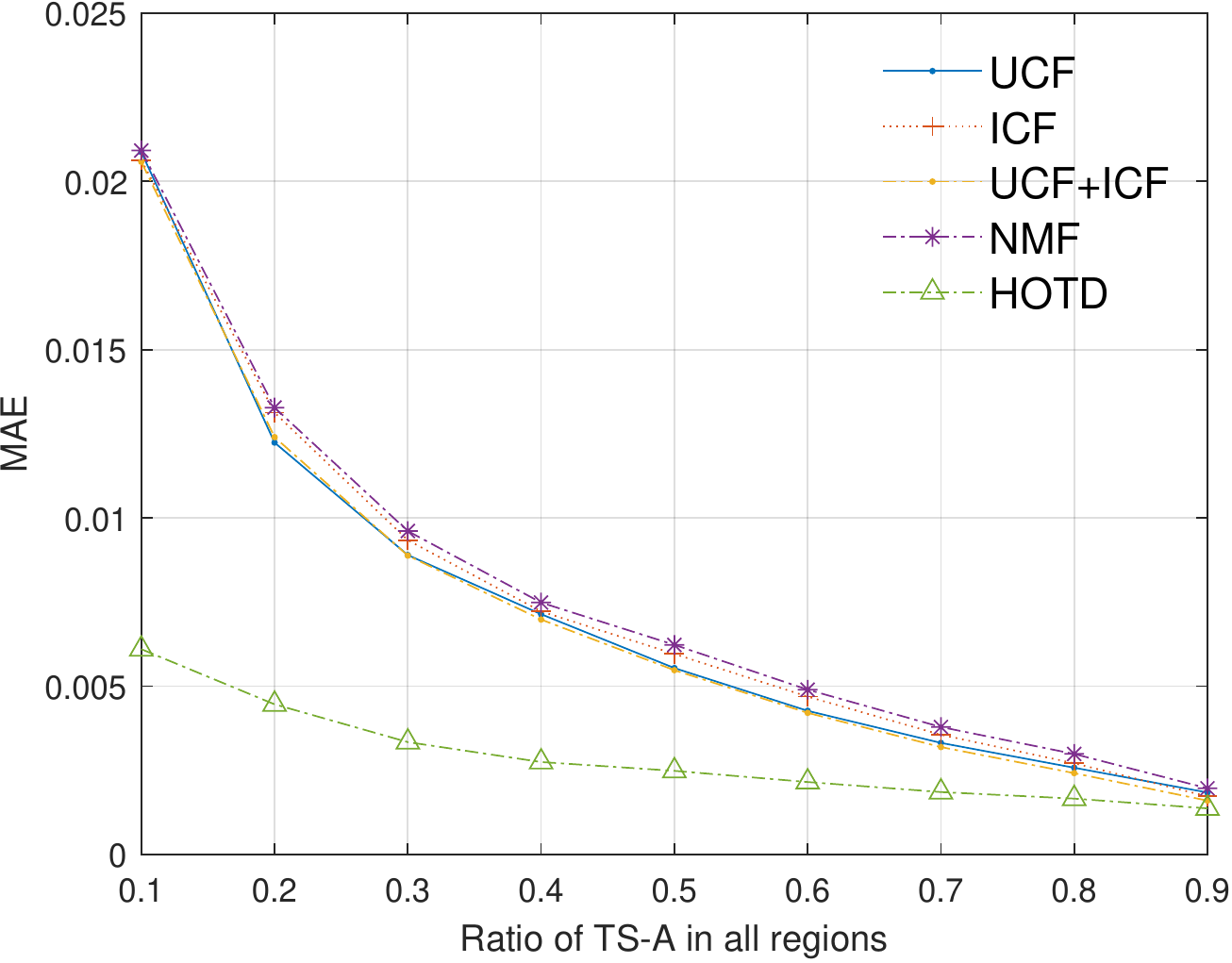}}
    \caption{Comparison of all data entry completion algorithms in QCB and RMDC with various proportion of TS-A}
    \label{fig:PerformanceComparison_QCB&RMDC}
\end{figure}

We further compare the performance of different data entry completion methods when adopting either QCB or RMDC as the TS-A selection method. The results in Figure~\ref{fig:PerformanceComparison_QCB&RMDC} show that HOTD outperforms other data entry completion methods in all settings\footnote{There are two curves covered by others}, which is consistent with the analysis in Section 4 that HOTD is able to better utilize the inter-disease correlations together with the spatial temporal correlations. 


\section{Related Work}

\hspace{1
em}{\itshape\textbf{Data Collections for PHM.}}
There are generally two ways for data collection in PHM. (a){\em The first one is to conduct population-based health surveys among a sample of residents within each target region,} in which healthcare administrators use questionnaires to obtain health-related information for each individual participant and then summarize to obtain the required statistics (e.g., obesity rate in different towns in south UK). There are generally two forms of health surveys: health interview surveys (HIS)~\cite{Blackwell2014Summary} and health examination surveys (HES)~\cite{Tolonen2018European}. HIS studies collect self-reported information via interviews or self-managed questionnaires, while HES collects more objective information via physical examinations and laboratory analysis. In order to minimize the risk of bias, the number of samples should be large enough to be representative to the population of interest, which makes the data collection process quite time-consuming and costly. (b){\em The other alternative for collecting data in PHM is through the integration of electronic health records (EHR) for clinic or hospital visits}~\cite{Kruse2018}. An important difference between EHR data and survey data is the fact that EHR data may have been originally collected for some other purposes rather than for data collection in PHM, which makes the data integration non-trivial due to the following reasons. First, the data access may be a barrier as some of data entries can be privacy-sensitive, and the data engineers need to do a lot of extra work (e.g., data anonymity operations)~\cite{De2015Using}. Second, the linkage is easiest in countries where personal identification numbers are used in all data sources. If only names, birth dates and addresses are available, it is often lead to problems to combine data sources as these identifiers may easily contain errors or change over time. Due to the above reasons, EHR data linkage still remains problematic, and a recent study on health data governance reported that only a minority of datasets with the capability to be linked were actually linked [10]. In summary, both two traditional data collection approaches for PHM are of high cost and time-consuming, which motivates us to propose the idea of CPH and evaluate its feasibility in this paper.

{\itshape\textbf{Population Health Prediction or Understanding.}}
Many existing studies have tried to understand the spread of infectious diseases and even predict their future trends. Some works utilized the social network structure and human mobility patterns to study the patterns of infectious diseases’ outbreak~\cite{Meyers2007Contact}, while some others utilize large amounts of users’ postings or tweets on social networks to predict the epidemic pattern on a large scale~\cite{Paul2011AAAI,Grover2014Prediction,Verma2017predicting}. In additional to the infectious disease, there are also literatures focusing on the prediction or understanding of chronic diseases. For example, the authors in~\cite{Yingzi2018IJCAI} utilized human mobility patterns of citizens to predict the evolution rate of several chronic diseases at the level of a city. The work~\cite{Mejova2015} employed both Foursquare and Instagram data to assess the relationship between fast food and obesity. Mason et al.~\cite{Mason2018Associations} found that people who lived far away from fast-food restaurants were more likely to have small waist circumference. The authors in~\cite{Trattner2017Monitoring} investigated the relationships between the online traces left behind by users of a large US online food community and the prevalence of obesity in 47 states and 311 counties in the US and discovered that higher fat and sugar content in bookmarked recipes was associated with higher rates of obesity. The authors of~\cite{Hasthanasombat2019Understanding} studied the effect of sport venue presence on the prevalence of anti-depressant prescriptions. While the above work utilizes external data source (e.g., social networks, mobilities, Point-of-Interests) to achieve the goal of population health prediction or understanding, this paper attempts to study the internal correlations (i.e., spatiotemporal correlations) for population health. Thus, we can see that the above works and our study are complementary to each other, and both can be jointly considered in the future to further understand population health.

{\itshape\textbf{Compressive Sensing and Its Applications.}}
Compressed sensing is a signal processing technique for efficiently acquiring and reconstructing a signal~\cite{Baraniuk2010Model,Baraniuk07}. A large number of applications based on compressive sensing have been proposed in recent years, such as network traffic reconstruction~\cite{Zhang2009Spatio}, environmental data recovery~\cite{Kong13,Quer12,Guestrin05}, road traffic monitoring~\cite{Zhu12}, urban crowd sensing~\cite{Leye15}, and face recognition on smartphones~\cite{Yiran14IEEE}. To the best of our knowledge, this is the first study attempting to leverage the idea of compressive sensing in the monitoring of population health and well-being.

\section{Summary and Discussions}
This paper proposed a novel cost-effective data collection approach for PHM named CPH and evaluated its feasibility in terms of three perspectives. In summary, the result of this study has led to the following findings. First, the spatiotemporal correlations of all studied chronic diseases’ morbidity rate generally do exist, but the correlations are very complicated as they are not linear, and the significance varies from one disease to another. This finding provides possibilities to further study the feasibility of CPH. Second, deploying the state-of-the-art missing data recovery techniques can achieve acceptable inference results for IF-A even when selecting a very small number of regions as TS-A. This result demonstrated that the inherent spatiotemporal correlations are strong enough to support the idea of CPH in real-world PHM tasks. Third, by comparing the performance of different TS-A selection approaches, our study further suggested that the optimized selection of TS-A can make CPH more cost-effective. 

With the above findings, we can see that the idea of CPH is generally feasible and has the potential to become a disruptive paradigm to achieve a more cost-effective PHM. However, this study should not end at this point, as it opens a door to a new direction which motivates the following future research efforts.

{\itshape\textbf{Generalizability Evaluation with More Datasets .}}
This paper investigates the feasibility of CPH by using one dataset. The research community including us needs to do more extensive analysis based on other datasets from different counties and cultures or containing different types of diseases. For example, it might be quite interesting to investigate the difference of spatiotemporal correlations between developed and developing countries.  Besides, the region of this study is set as a ward level, but it is also interesting to explore the correlations when the granularity of each region changes to other levels (e.g., a town or a county). Furthermore, this paper only focuses on the chronic disease due to the limitation of datasets. In future work, we also attempt to study the same problem on infectious diseases and compare the inter-disease difference in spatiotemporal correlations with deeper insights.

{\itshape\textbf{More Sophisticated Algorithms.}}
The major contribution of this paper is to propose the idea of CPH and evaluate its feasibility, but we do not claim that the adopted algorithms are our technical contributions. In fact, there are still large room to improve the key algorithms within the general framework of CPH, including both the missing data completion and the selection of TS-A. First, we can build a hybrid missing entries recovery model by integrating different state-of-the-art algorithms to handle different data missing patterns (e.g., block missing, different proportion of missing data, and so on). Second, we can also leverage more sophisticated machine learning algorithms (e.g., reinforcement learning and active learning) to further improve the performance of TS-A selection module.

{\itshape\textbf{Factors Accounting for the Spatiotemporal Correlations.}}
Now that we have proved the existence of the spatiotemporal correlations, it is interesting to explore what factors account this correlation (for example, age, gender, population migration, and so on), which are insightful to improve the depth of this work. However, the major purpose of this paper is to evaluate the feasibility of the compressive approach in reducing the data collection cost of PHM by answering three research questions, rather than studying the factors behind the correlations. Actually, there is well-studied sub-field of public health called “Spatial Epidemiology”~\cite{Lawson2016Handbook}, which studies the spatial determinants to cause the inter-region health inequality. In the future work, we aim to combine our study of CPH and state-of-the-art research work on Spatial Epidemiology to move this work to a deeper level. 

{\itshape\textbf{Combination with Other Correlations.}}
The basic intuition of CPH is to leverage the inherent data correlations to do inference thus reducing the cost in data collection. This paper mainly investigates the spatiotemporal correlations and explore if they can be used to support CPH. However, there are also other correlations that we can exploit in the future to further reduce the collection cost or improve the inference accuracy. First, Multi-source urban big data have become widely available, e.g., population density, education and economic status, age distribution, human mobility, and distributions of POIs (Point-of-Interests), air quality measures, and so on. Some of these data sources have relationships with the population health status. For example, previous studies found strong correlations between the density of fast food venues and BMI (Body Mass Index)~\cite{Mason2018Associations}, while studies such as~\cite{Yingzi2018IJCAI} indicates that human’s mobility patterns are correlated to the chronic disease morbidity. Second, we can also use inter-disease correlations to improve the inference. There is a phenomenon called multi-morbidity~\cite{Barnett2012}, which is commonly defined as the presence of two or more chronic medical conditions. With the multi-morbidity in mind, for example, the regions with higher rate of obesity are more likely to have higher rate of heart attack and cancers.

{\itshape\textbf{More Practical Factors.}}
Although this study demonstrates the feasibility of CPH and its potential impact, there are still a lot of important factors which should be further taken into account if we attempt to build a real-world CPH system. For example, in our second and third study, we demonstrate that with proper data entry recovery algorithms and TS-A selection methods, CPH can achieve an acceptable accuracy with a certain fixed number of TS-A. However, in practical applications, the accuracy requirement can vary for different diseases or countries. Thus, it would be important to study how to select a minimum number of TS-A while ensuring a certain level of accuracy, and evaluate how much cost can be saved. Second, the cost of data collection for different regions may not be the same, which should be further taken into the optimization process in the future research efforts. For example, it will be more costly for regions with larger populations to get a representative group of people for surveys or lab tests.

\section{Conclusions}
Motivated by the requirement of cost reduction in the monitoring of population health, this paper proposed a cost-effective approach called CPH for population health monitoring, whose main idea is to select a subset of regions to conduct traditional health data collection, while leveraging inherent data correlations to do data inference for the rest of un-collected regions. To verify whether this idea is feasible, this paper conducted an in-depth study based on spatiotemporal morbidity rate data of 18 types of chronic diseases in more than 500 ward-level regions in London, UK. By a data-driven study from different views, the results revealed the existence of significant spatiotemporal correlations, and demonstrated that state-of-the-art data recovery methods can utilize these correlations to do data inference accurately with a relatively small number of samples. We also suggested that more sophisticated selections for the regions of population health data collection can further optimize the cost and quality of monitoring.

\bibliographystyle{ACM-Reference-Format}
\bibliography{sample-base}

\end{document}